\documentclass[twocolumn]{aastex63}

\usepackage[utf8]{inputenc}

\usepackage{amsmath}
\usepackage{enumitem}
\newlist{inlinelist}{enumerate*}{1}
\setlist[inlinelist]{label=(\roman*)}

\newcommand{\solarmass}{\ensuremath{\mathrm{M}_{\odot}}}

\newcommand{\solarmet}{\ensuremath{\mathrm{Z}_{\odot}}}
\newcommand{\Msun}{\solarmass}

\newcommand{\Zsun}{\solarmet}

\newcommand{\Mdust}{\ensuremath{M_{\mathrm{dust}}}}

\newcommand{\Mmol}{\ensuremath{M_{\mathrm{mol}}}}

\newcommand{\HH}{\ensuremath{\mathrm{H}_{2}}}

\newcommand{\rhomol}{\ensuremath{\rho_{\mathrm{mol}}}}

\newcommand{\logOH}{\ensuremath{12 + \log{(\mathrm{O}/\mathrm{H})}}}

\newcommand{\mugas}{\ensuremath{\mu_{\mathrm{gas}}}}
\newcommand{\fgas}{\ensuremath{f_{\mathrm{gas}}}}
\newcommand{\aco}{\ensuremath{\alpha_{\mathrm{CO}}}}

\newcommand{\Tex}{\ensuremath{T_{\mathrm{ex}}}}
\newcommand{\Tdust}{\ensuremath{T_{\mathrm{dust}}}}
\newcommand{\gdr}{\ensuremath{\delta_{\mathrm{GDR}}}}

\newcommand{\avg}[1]{\ensuremath{\langle #1 \rangle}}

\newcommand{\Vmag}{\ensuremath{V_{\mathrm{F606W}}}}
\newcommand{\Imag}{\ensuremath{i_{\mathrm{F775W}}}}
\newcommand{\Hmag}{\ensuremath{H_{\mathrm{F160W}}}}

\newcommand{\Lyalpha}{\ensuremath{\mathrm{Ly}\alpha\,\lambda 1216}}

\newcommand{\Civ}{\ensuremath{\mathrm{C\,\textsc{iv}}\,\lambda\lambda 1548, 1551}}
\newcommand{\Heii}{\ensuremath{\mathrm{He\,\textsc{ii}}\,\lambda 1640}}
\newcommand{\Oiiiuv}{\ensuremath{\mathrm{O\,\textsc{iii}]}\,\lambda\lambda 1661, 1666}}

\newcommand{\Siiii}{\ensuremath{\mathrm{S\,\textsc{iii}]}\,\lambda\lambda 1883, 1892}}
\newcommand{\Ciii}{\ensuremath{\mathrm{C\,\textsc{iii}]}\,\lambda\lambda 1907, 1909}}

\newcommand{\Siiia}{\ensuremath{\mathrm{Si\,\textsc{ii}}\,\lambda 1260}}
\newcommand{\Oi}{\ensuremath{\mathrm{O\,\textsc{ii}}\,\lambda 1302}}
\newcommand{\Siiib}{\ensuremath{\mathrm{Si\,\textsc{ii}}\,\lambda 1304}}
\newcommand{\Cii}{\ensuremath{\mathrm{C\,\textsc{ii}}\,\lambda 1335}}
\newcommand{\Siiv}{\ensuremath{\mathrm{Si\,\textsc{iv}}\,\lambda\lambda 1394,1403}}
\newcommand{\Feii}{\ensuremath{\mathrm{Fe\,\textsc{ii}}\,\lambda\lambda 1608,1611}}
\newcommand{\Alii}{\ensuremath{\mathrm{Al\,\textsc{ii}}\,\lambda 1671}}
\newcommand{\Aliii}{\ensuremath{\mathrm{Al\,\textsc{ii}}\,\lambda\lambda 1854,1862}}

\newcommand{\Oiialt}{\ensuremath{\mathrm{[O\,\textsc{ii}]}\,\lambda 3727}}

\newcommand{\Oiii}{\ensuremath{\mathrm{[O\,\textsc{iii}]}\,\lambda\lambda 4960,5008}}
\newcommand{\Oiiia}{\ensuremath{\mathrm{[O\,\textsc{iii}]}\,\lambda 4960}}
\newcommand{\Oiiib}{\ensuremath{\mathrm{[O\,\textsc{iii}]}\,\lambda 5008}}

\newcommand{\Hbeta}{\ensuremath{\mathrm{H}\beta\,\lambda 4863}}

\newcommand{\Lya}{\ensuremath{\mathrm{Ly}\alpha}}

\newcommand{\Hb}{\ensuremath{\mathrm{H}\beta}}

\newcommand{\HI}{\ensuremath{\mathrm{H\,\textsc{i}}}}
\newcommand{\HII}{\ensuremath{\mathrm{H\,\textsc{ii}}}}

\newcommand{\OIII}{\ensuremath{\mathrm{[O\,\textsc{iii}]}}}

\newcommand{\CI}{\ensuremath{\mathrm{[C\, \textsc{i}]}}}
\newcommand{\CII}{\ensuremath{\mathrm{[C\, \textsc{ii}]}}}

\newcommand{\CIthreePone}{\ensuremath{\mathrm{[C\, \textsc{i}]~{^{3}P_{1}} \rightarrow {^{3}P_{0}}}}}

\newcommand{\OIIIfslb}{\ensuremath{\mathrm{[O\, \textsc{iii}]\,\lambda88\,\micron}}}
\newcommand{\CIIfsl}{\ensuremath{\mathrm{[C\, \textsc{ii}]\,\lambda158\,\micron}}}
\newcommand{\CIIonehalfP}{\ensuremath{\mathrm{[C\, \textsc{ii}]~{^{3}P_{3/2}} \rightarrow {^{3}P_{1/2}}}}}

\newcommand{\Nsample}{\ensuremath{168}}

\newcommand{\Nobj}{\ensuremath{24}}
\newcommand{\Nlae}{\ensuremath{20}}

\newcommand{\zavg}{\ensuremath{3.45}}
\newcommand{\Mmed}{\ensuremath{10^{9.1}\,\mathrm{M}_{\odot}}}
\newcommand{\SFRmed}{\ensuremath{10\,\mathrm{M}_{\odot}\,\mathrm{yr}^{-1}}}

\newcommand{\logOHmed}{\ensuremath{7.7_{-0.2}^{+0.3}}}
\newcommand{\logOHavg}{\ensuremath{7.8 \pm 0.2}}

\newcommand{\acoavg}{\ensuremath{50}}%
\newcommand{\gdravglin}{\ensuremath{550}}%
\newcommand{\gdravgbpl}{\ensuremath{3300}}%

\newcommand{\acomin}{\ensuremath{10}}
\newcommand{\gdrmin}{\ensuremath{1200}}

\newcommand{\LLyaavg}{\ensuremath{10^{42}\,\mathrm{erg}\,\mathrm{s}^{-1}}}
\newcommand{\EWavg}{\ensuremath{20}}

\newcommand{\logOIIIHbKMOSstack}{\ensuremath{0.9 \pm 0.3}}
\newcommand{\logOHKMOSstack}{\ensuremath{7.6-8.0}}

\newcommand{\deltavlya}{\ensuremath{346}}
\newcommand{\deltavstack}{\ensuremath{300}}
\newcommand{\cofourstackflux}{\ensuremath{-3.2 \pm 4.2}}%
\newcommand{\cofourstacklimit}{\ensuremath{12.7}}%
\newcommand{\cofourstacknum}{\ensuremath{23}}
\newcommand{\cofourstacklimitLp}{\ensuremath{4.0 \times 10^{8}}} %

\newcommand{\cistackflux}{\ensuremath{4.9 \pm 6.8}}%
\newcommand{\cistacklimit}{\ensuremath{20.5}}%
\newcommand{\cistacknum}{\ensuremath{16}}
\newcommand{\cistacklimitLp}{\ensuremath{5.6 \times 10^{8}}} %

\newcommand{\coninestackflux}{\ensuremath{33.3 \pm 12.7}}%
\newcommand{\coninestacklimit}{\ensuremath{38.1}}%
\newcommand{\coninestacknum}{\ensuremath{23}}
\newcommand{\coninestacklimitLp}{\ensuremath{2.3 \times 10^{8}}} %

\newcommand{\cotenstackflux}{\ensuremath{7.4 \pm 16.6}}%
\newcommand{\cotenstacklimit}{\ensuremath{49.8}}%
\newcommand{\cotenstacknum}{\ensuremath{16}}
\newcommand{\cotenstacklimitLp}{\ensuremath{2.5 \times 10^{8}}} %

\newcommand{\cofourstackfluxlae}{\ensuremath{-7.3 \pm 5.0}}%
\newcommand{\cofourstacklimitlae}{\ensuremath{15.0}}%
\newcommand{\cofourstacknumlae}{\ensuremath{18}}
\newcommand{\cofourstacklimitLplae}{\ensuremath{4.7 \times 10^{8}}} %

\newcommand{\cistackfluxlae}{\ensuremath{11.0 \pm 7.9}}%
\newcommand{\cistacklimitlae}{\ensuremath{23.8}}%
\newcommand{\cistacknumlae}{\ensuremath{12}}
\newcommand{\cistacklimitLplae}{\ensuremath{6.5 \times 10^{8}}} %

\newcommand{\coninestackfluxlae}{\ensuremath{39.5 \pm 14.0}}%
\newcommand{\coninestacklimitlae}{\ensuremath{42.2}}%
\newcommand{\coninestacknumlae}{\ensuremath{19}}
\newcommand{\coninestacklimitLplae}{\ensuremath{2.6 \times 10^{8}}} %

\newcommand{\cotenstackfluxlae}{\ensuremath{16.2 \pm 18.8}}%
\newcommand{\cotenstacklimitlae}{\ensuremath{56.5}}%
\newcommand{\cotenstacknumlae}{\ensuremath{13}}
\newcommand{\cotenstacklimitLplae}{\ensuremath{2.8 \times 10^{8}}} %

\newcommand{\duststack}{\ensuremath{1 \pm 3\,\mu\mathrm{Jy}}}
\newcommand{\dustlimit}{\ensuremath{9\,\mu\mathrm{Jy}}}

\newcommand{\dustlimitThreemm}{\ensuremath{1.2\,\mu\mathrm{Jy}}}

\newcommand{\colimitmmolsolar}{\ensuremath{5.74}}  %
\newcommand{\colimitmusolar}{\ensuremath{4.56}}  %
\newcommand{\cilimitmmolsolar}{\ensuremath{7.91}}  %
\newcommand{\cilimitmusolar}{\ensuremath{6.29}}  %
\newcommand{\dustlimitmmolsolar}{\ensuremath{1.09}}  %
\newcommand{\dustlimitmusolar}{\ensuremath{0.87}}  %

\newcommand{\colimitmmolsubsolar}{\ensuremath{65.9}}  %
\newcommand{\colimitmusubsolar}{\ensuremath{52.3}}  %
\newcommand{\cilimitmmolsubsolar}{\ensuremath{63.3}}  %
\newcommand{\cilimitmusubsolar}{\ensuremath{50.3}}  %
\newcommand{\dustlimitmmolsubsolarlin}{\ensuremath{5.99}}  %
\newcommand{\dustlimitmusubsolarlin}{\ensuremath{4.76}}  %
\newcommand{\dustlimitmmolsubsolarbpl}{\ensuremath{36.0}}  %
\newcommand{\dustlimitmusubsolarbpl}{\ensuremath{28.6}}  %

\newcommand{\zs}[1]{\ensuremath{z_{\mathrm{#1}}}}

\newcommand{\zLya}{\ensuremath{\zs{\Lya}^{\mathrm{red}}}}

\newcommand{\zsys}{\zs{sys}}

\newcommand{\EWLya}{\ensuremath{\mathrm{EW}^{0}_{\Lya}}}

\received{23 October 2020}
\revised{12 April 2021}
\accepted{13 May 2021}

\submitjournal{ApJ}

\shorttitle{Molecular gas in star-forming galaxies at $\avg{z}=\zavg$}
\shortauthors{Boogaard et al.}

\begin{document}

\title{Measuring the average molecular gas content of star-forming galaxies at
  $z=3-4$}

\correspondingauthor{Leindert Boogaard}
\email{boogaard@strw.leidenuniv.nl}

\author[0000-0002-3952-8588]{Leindert A. Boogaard} \affil{Leiden Observatory,
  Leiden University, PO Box 9513, NL-2300 RA Leiden, The Netherlands}

\author[0000-0002-4989-2471]{Rychard J. Bouwens} \affil{Leiden Observatory,
  Leiden University, PO Box 9513, NL-2300 RA Leiden, The Netherlands}

\author[0000-0001-9585-1462]{Dominik Riechers} \affil{Cornell University, 220
  Space Sciences Building, Ithaca, NY 14853, USA} \affil{Max Planck Institute
  f\"ur Astronomie, K\"onigstuhl 17, 69117 Heidelberg, Germany}

\author[0000-0001-5434-5942]{Paul van der Werf} \affil{Leiden Observatory,
Leiden University, PO Box 9513, NL-2300 RA Leiden, The Netherlands}

\author{Roland Bacon} \affil{Univ. Lyon 1, ENS de Lyon, CNRS, Centre de
Recherche Astrophysique de Lyon (CRAL) UMR5574, 69230 Saint-Genis-Laval,
France}

\author[0000-0003-2871-127X]{Jorryt Matthee} \affil{Department of Physics, ETH Zurich,
Wolfgang-Pauli-Strasse 27, 8093, Zurich, Switzerland}

\author[0000-0001-7768-5309]{Mauro Stefanon} \affil{Leiden Observatory,
  Leiden University, PO Box 9513, NL-2300 RA Leiden, The Netherlands}

\author[0000-0001-6865-2871]{Anna Feltre} \affil{INAF-Osservatorio di Astrofisica e Scienza dello
  Spazio, via Gobetti 93/3, I-40129, Bologna, Italy}

\author[0000-0003-0695-4414]{Michael Maseda} \affil{Leiden Observatory,
  Leiden University, PO Box 9513, NL-2300 RA Leiden, The Netherlands}

\author{Hanae Inami} \affil{Hiroshima Astrophysical Science Center, Hiroshima
  University, 1-3-1 Kagamiyama, Higashi-Hiroshima, Hiroshima, 739-8526}

\author[0000-0002-6290-3198]{Manuel Aravena} \affil{N\'ucleo de Astronom\'ia
de la Facultad de Ingenier\'ia y Ciencias, Universidad Diego Portales,
Av. Ej\'ercito Libertador 441, Santiago, Chile}

\author[0000-0003-4359-8797]{Jarle Brinchmann} \affil{Leiden Observatory, Leiden University, PO Box
  9513, NL-2300 RA Leiden, The Netherlands} \affil{Instituto de
  Astrof{\'\i}sica e Ci{\^e}ncias do Espa\c{c}o, Universidade do Porto, CAUP, Rua
  das Estrelas, PT4150-762 Porto, Portugal}

\author[0000-0001-6647-3861]{Chris Carilli} \affil{National Radio Astronomy
  Observatory, Pete V. Domenici Array Science Center, P.O. Box O, Socorro, NM
  87801, USA} \affil{Battcock Centre for Experimental Astrophysics, Cavendish
  Laboratory, Cambridge CB3 0HE, UK}

\author[0000-0003-0275-938X]{Thierry Contini} \affil{Institut de Recherche en
  Astrophysique et Plan\'etologie (IRAP), Universit\'e de Toulouse, CNRS, UPS,
  31400 Toulouse, France}

\author[0000-0002-2662-8803]{Roberto Decarli} \affil{INAF-Osservatorio di
Astrofisica e Scienza dello Spazio, via Gobetti 93/3, I-40129, Bologna,
Italy}

\author[0000-0003-3926-1411]{Jorge Gonz\'alez-L\'opez} \affil{N\'ucleo de
  Astronom\'ia de la Facultad de Ingenier\'ia y Ciencias, Universidad Diego
  Portales, Av. Ej\'ercito Libertador 441, Santiago, Chile} \affil{Instituto de
  Astrof\'{\i}sica, Facultad de F\'{\i}sica, Pontificia Universidad Cat\'olica
  de Chile Av. Vicu\~na Mackenna 4860, 782-0436 Macul, Santiago, Chile}

\author[0000-0003-2804-0648]{Themiya Nanayakkara} \affil{Centre for
Astrophysics and Supercomputing, Swinburne University of Technology,
Hawthorn, VIC 3122, Australia}

\author[0000-0003-4793-7880]{Fabian Walter} \affil{Max Planck Institute f\"ur
  Astronomie, K\"onigstuhl 17, 69117 Heidelberg, Germany} \affil{National Radio
  Astronomy Observatory, Pete V. Domenici Array Science Center, P.O. Box O,
  Socorro, NM 87801, USA}

\begin{abstract}
  We study the molecular gas content of \Nobj\ star-forming galaxies at
  $z=3-4$, with a median stellar mass of \Mmed, from the MUSE \emph{Hubble}
  Ultra Deep Field (HUDF) Survey.  Selected by their \Lyalpha-emission and
  \Hmag-band magnitude, the galaxies show an average
  $\avg{\EWLya} \approx \EWavg$\,\AA, below the typical selection threshold for
  Lyman Alpha Emitters ($\EWLya > 25$\,\AA), and a rest-frame UV spectrum
  similar to Lyman Break Galaxies.  We use rest-frame optical spectroscopy from
  KMOS and MOSFIRE, and the UV features observed with MUSE, to determine the
  systemic redshifts, which are offset from \Lya\ by
  $\avg{\Delta v(\Lya)} = \deltavlya$~km\,s$^{-1}$, with a 100 to
  600\,km\,s$^{-1}$ range.  Stacking $^{12}$CO~$J=4\rightarrow3$ and
  \CIthreePone\ (and higher-$J$ CO lines) from the ALMA Spectroscopic Survey of
  the HUDF (ASPECS), we determine $3\sigma$ upper limits on the line
  luminosities of \cofourstacklimitLp~K\,km\,s$^{-1}$pc$^{2}$ and
  \cistacklimitLp~K\,km\,s$^{-1}$pc$^{2}$, respectively (for a
  \deltavstack~km\,s$^{-1}$ linewidth).  Stacking the 1.2\,mm and 3\,mm dust
  continuum flux densities, we find a $3\sigma$ upper limits of $\dustlimit$
  and $\dustlimitThreemm$, respectively.  The inferred gas fractions, under the
  assumption of a `Galactic' CO-to-H$_{2}$ conversion factor and gas-to-dust
  ratio, are in tension with previously determined scaling relations.  This
  implies a substantially higher $\aco \ge \acomin$ and $\gdr \ge \gdrmin$,
  consistent with the sub-solar metallicity estimated for these galaxies
  ($\logOH \approx \logOHavg$).  The low metallicity of $z\ge3$ star-forming
  galaxies may thus make it very challenging to unveil their cold gas through
  CO or dust emission, warranting further exploration of alternative tracers,
  such as \CII.
\end{abstract}


\section{Introduction}
\label{sec:introduction}
The recent decade has witnessed a tremendous advance in the characterization of
the cold molecular gas content of star forming galaxies at $z >1$.  Evidence is
now mounting that the cold gas fraction of massive star-forming galaxies
strongly increases up to at least $z\approx 3$ \citep[e.g.,][]{Tacconi2010,
  Tacconi2013, Tacconi2018, Genzel2010, Genzel2015, Geach2011,
  Dessauges-Zavadsky2015, Dessauges-Zavadsky2020, Aravena2019, Aravena2020,
  Tacconi2020}.  As the cold H$_{2}$ gas itself is radiatively dark, the
molecular gas has to be traced by the emission from the ground-state rotational
transition of Carbon Monoxide ($^{12}$CO, hereafter CO), or other tracers such
as the emission from neutral atomic carbon (\CI) or the long-wavelength dust
continuum.  Yet, observations of CO in (optically selected) star-forming
galaxies at $z>3$ remain challenging and have been limited to massive Lyman
Break- or main sequence-selected galaxies and/or strongly lensed systems, with
known redshifts \citep{Baker2004, Coppin2007, Riechers2010, Magdis2012,
  Magdis2017, Tan2013, Livermore2012, Saintonge2013, Dessauges-Zavadsky2015,
  Dessauges-Zavadsky2017, Pavesi2019, Cassata2020a}, sometimes serendipitously
detected and only identified as such \emph{a posteriori} \citep{Gowardhan2019}.

The Atacama Large Millimeter Array Large Program ASPECS (The ALMA Spectroscopic
Survey in the \emph{Hubble} Ultra Deep Field (HUDF); \citealt{Walter2016,
  Decarli2019}) provides a unique opportunity to study the gas content of star
forming galaxies at $z\ge3$.  ASPECS consists of spectral scans in ALMA Band 3
(85--115\,GHz) and 6 (212--272\,GHz), probing molecular gas and dust in
galaxies without any target preselection.  These data unveil emission from CO,
\CI\ and dust-continuum in several star-forming galaxies at $z=1-4$
\citep{Gonzalez-Lopez2019, Gonzalez-Lopez2020, Boogaard2019, Boogaard2020},
providing key constraints on the empirical scaling relations describing the
evolution of the gas and dust content in galaxies \citep{Aravena2019,
  Aravena2020}, the evolution of the cosmic molecular gas
density \citep{Decarli2019, Decarli2020} and the baryon cycle
\citep{Walter2020}.

Key to the exploration of the ASPECS data are the large number of spectroscopic
redshifts provided by the Multi Unit Spectroscopic Explorer (MUSE) HUDF Survey
\citep{Bacon2017}.  Through its unparalleled sensitivity for faint emission
lines, MUSE is very efficient in detecting galaxies at $z \geq 3$, where the
bright \HI\ \Lyalpha\ line enters the integral-field spectrograph
($4750-9300$\AA; $\lambda/\Delta\lambda \approx 3000$; \citealt{Inami2017}),
probing the faint end of the \Lya\ luminosity function down to below
$0.1\,L_{\rm Lya}^{*}$ \citep{Drake2017}.

Exploiting the large number of redshifts from MUSE, we can push the gas
mass-sensitivity of ASPECS at $z\geq 3$ to its limits through stacking (in
particular, CO~$J=4 \rightarrow 3$ becomes accessible at $z \geq 3.0115$).
Indeed, \cite{Inami2020} have shown that at lower redshifts, $z=1-2$, we can
recover CO emission below the formal sensitivity threshold of ASPECS, by
stacking on the accurate systemic redshifts from MUSE.

However, the MUSE redshifts at $z\geq3$ obtained from \Lya\ cannot be used for
stacking.  Because \Lya\ is a resonant transition, the photons are easily
scattered by the neutral gas in- and surrounding a galaxy, shifting the peak of
the emission away from the systemic velocity by several hundred km~s$^{-1}$
\citep[e.g.,][]{Shapley2003, Verhamme2018, Muzahid2020}.  This means that the
line emission tracing the molecular gas could be completely washed out by the
noise if non-systemic \Lya-redshifts are used for stacking.

Fortunately, because we have approximate redshifts from Ly$\alpha$, these
targets can be effectively followed-up simultaneously with multi-object,
near-infrared spectrographs, such as the K-band Multi Object Spectrograph
(KMOS) at the Very Large Telescope and the Multi-Object Spectrometer For
Infra-Red Exploration (MOSFIRE) at the Keck Observatory.  These data can
constrain the systemic redshift by targeting the rest-frame optical lines such
as \Hbeta\ and \Oiii\ in the \emph{K}-band (out to $z=3.812$).

In this paper, we constrain the molecular gas content of star-forming galaxies
at $3.0115 < z < 3.812$, by stacking their molecular gas signal through the
outlined three step process. We 1) identify the galaxies from MUSE and 2)
determine their systemic redshifts through rest-frame NIR spectroscopy with
KMOS/MOSFIRE (as well as rest-UV features from MUSE;
\autoref{sec:observations}).  We then 3) turn to ALMA to stack the CO and \CI\
signal from the ASPECS data, as well as the 1.2\,mm dust continuum
(\autoref{sec:results}).  We do not detect any (line) emission in the stacks
(at the $3\sigma$ level) and discuss the implications of this non-detection on
metallicity, the CO-to-H$_{2}$ conversion factor (\aco) and gas-to-dust ratio
(\gdr) in \autoref{sec:disc}.  The results highlight that the metallicity
evolution of star-forming galaxies makes it increasingly challenging to infer
the molecular gas content at higher redshifts, which warrants the further
theoretical and observational exploration of alternative tracers, in particular
the \CIIfsl\ line.

Throughout this paper, we report wavelengths in vacuo and magnitudes in the AB
system \citep{Oke1983}, and adopt a \cite{Chabrier2003} initial mass function.
We use $\log$ to denote $\log_{10}$ and $\ln$ for the natural logarithm.  We
adopt a concordance cosmology with $H_0 = 70$\,km\,s$^{-1}$\,Mpc$^{-1}$,
$\Omega_{m} = 0.3$ and $\Omega_{\Lambda} = 0.7$, in good agreement with the
measurements from \cite{PlanckCollaboration2015}.

\section{Observations and sample selection}
\label{sec:observations}
\begin{figure}[t]
  \includegraphics[width=\columnwidth]{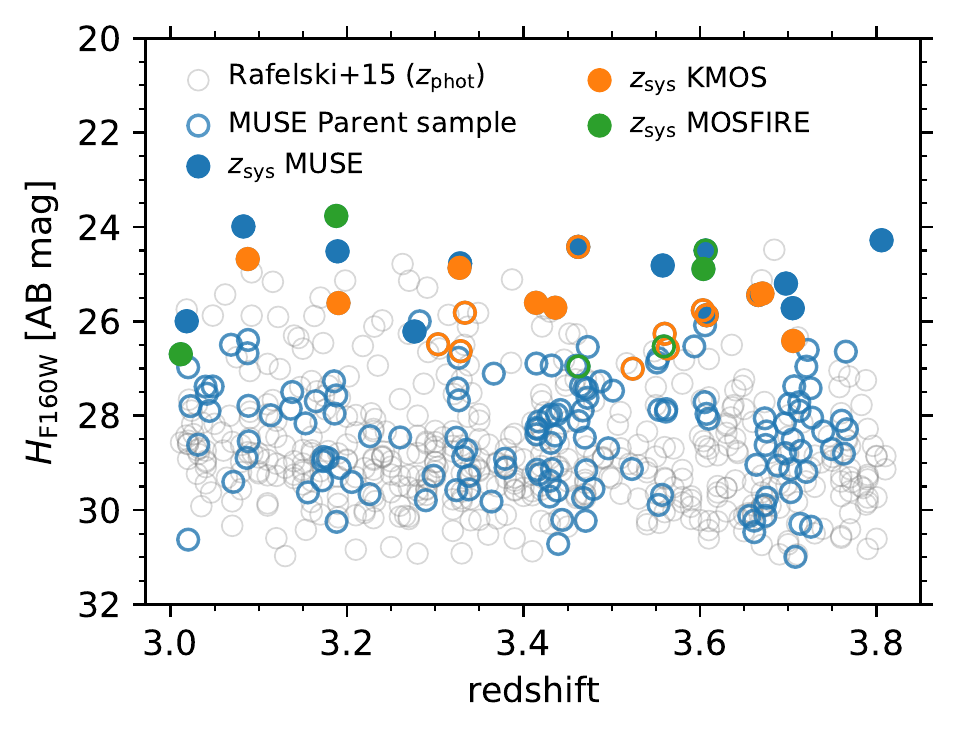}
  \caption{The \Hmag-band magnitude of the galaxies considered in this paper as
    a function of redshift.  The open circles show the parent sample of all
    galaxies with $3.0115\le z < 3.812$ within the ASPECS field ($\ge40\%$ of
    the primary beam peak sensitivity) that have a MUSE redshift (from
    \Lyalpha) in blue, while the gray circles show all galaxies in the same
    field with a photometric redshift \citep{Rafelski2015}.  Galaxies targeted
    for KMOS or MOSFIRE follow-up are shown in orange and green, respectively.
    The filled circles show galaxies for which we obtained a systemic redshift
    measurement from MUSE (blue), KMOS (orange) or MOSFIRE (green).
    \label{fig:z-f160}}
\end{figure}
\begin{figure*}[t]
  \centering
  \includegraphics[width=\textwidth]{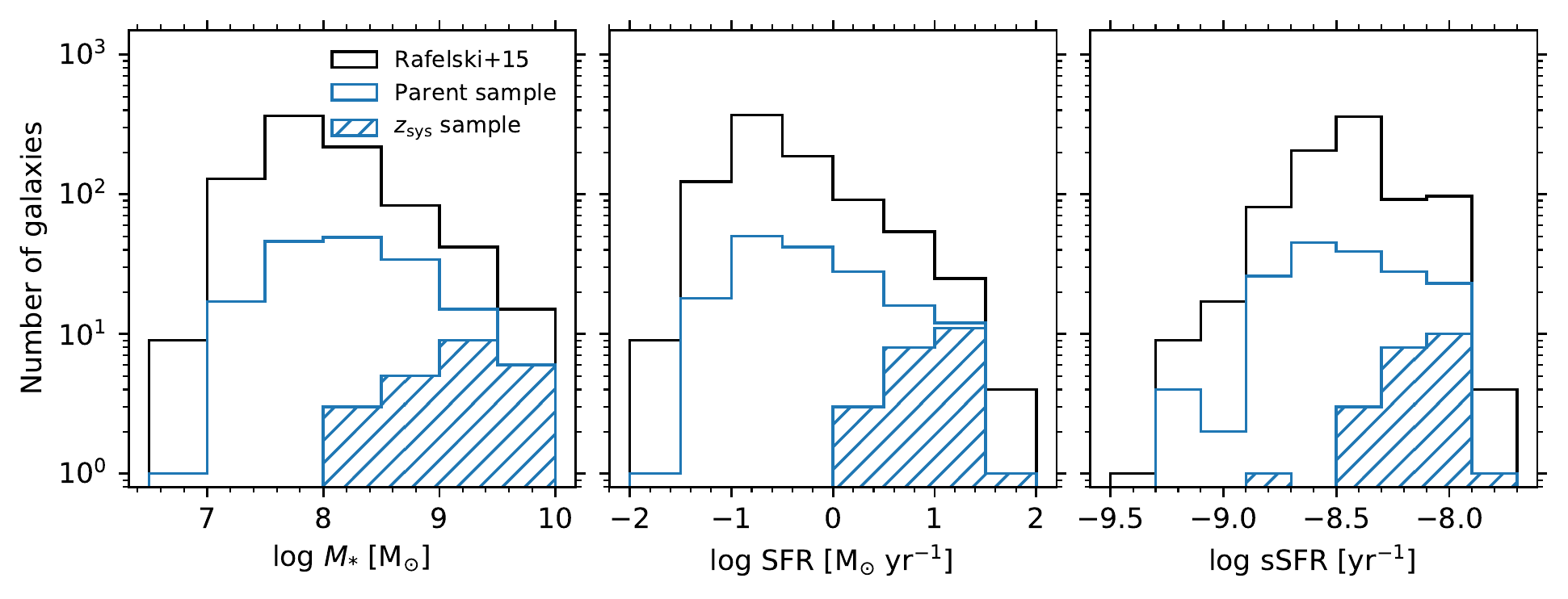}
  \caption{Histograms of the stellar mass ($M_{*}$), star formation rate (SFR)
    and specific SFR ($\mathrm{sSFR} = \mathrm{SFR}/M_{*}$) of the galaxies for
    which we determine a systemic redshift (\zsys), compared to the parent
    sample from MUSE at $3.0115\le z < 3.812$ and all galaxies from the
    photometric redshift catalog by \cite{Rafelski2015}.\label{fig:hist}}
\end{figure*}

\subsection{Parent sample selection and physical properties}
\label{sec:sample-selection}
We construct a parent sample of galaxies from the MUSE HUDF Survey Data Release
2 catalog,\footnote{DR2 v0.1; R. Bacon et al., in prep.} which is an
updated and revised version of the DR1 catalog \citep{Bacon2017, Inami2017}.
In short, the catalog contains both emission line-selected sources \citep[from
ORIGIN;][]{Mary2020} and continuum-selected sources (from the \emph{Hubble
  Space Telescope (HST)} catalog by \citealt{Rafelski2015}) for which the
redshifts are determined automatically.  These sources have subsequently been
verified by several independent groups of experts that inspect the redshift,
the multiwavelength counterpart associations, and assign a confidence flag
(ZCONF; where confidence $\ge 2$ implies a secure redshift, determined by at
least two spectral features).  Specifically, we use the following criteria:
\begin{itemize}
\item Select all objects with $3.0115 < z <3.812$ and $\mathrm{ZCONF} \ge 2$,
  that have a \emph{HST} counterpart in the \cite{Rafelski2015} catalog.
\item Restrict to objects that lie within the 4.55 arcmin$^{2}$ region of the
  ASPECS Band 3 mosaic where the sensitivity is $\ge 40\%$ of the primary beam
  peak sensitivity at 99.5\,GHz.\footnote{This area fully encompasses the
    ASPECS Band 6 mosaic.}
\item Remove three X-ray detected sources that are classified as having an
  active galactic nucleus (AGN; MUSE-1051, MUSE-1056 and MUSE-6672), based on the
  \emph{Chandra} 7\,MS data \citep{Luo2017}.
\end{itemize}
There are a total of \Nsample\ galaxies in the parent sample constructed this
way.  The \Hmag\ magnitude of the parent sample is shown as a function of
redshift in \autoref{fig:z-f160}.  Because of the sensitivity of MUSE to
faint emission line sources, it consists almost exclusively of galaxies that
are selected by their \Lyalpha-emission.  Only six galaxies are not marked as
such: one is MUSE-50, which does show double peaked \Lya-emission on top of
strong \Lya-absorption, as well as strong UV lines.  The other five indeed show
little \Lya-emission: one is a faint \Civ-only-emitter, while the other four
have bright enough UV continuum to have their systemic redshifts determined
from absorption lines (see \autoref{sec:muse}).

Because the \Lya\ emission may peak in the halo of a galaxy, the association of
a MUSE source with an \emph{HST} counterpart can be ambiguous and is typically
resolved during the redshift determination process.  The associations adopted
here are listed in \autoref{tab:sources} and are in all cases supported by a
second tracer of the systemic redshift.  In the case of MUSE-6518, the
photometry is completely blended with a $z=0.83$ foreground object and we do
not use it to obtain physical properties.

We determine a stellar mass ($M_{*}$) and star-formation rate (SFR) for all
galaxies in the parent sample by fitting eleven bands of \emph{HST}
\citep{Rafelski2015} and four bands of \emph{Spitzer}/IRAC photometry, using
the high-$z$ extension of the spectral energy distribution fitting code
\textsc{Magphys} \citep{DaCunha2008,DaCunha2015}.  As in \citet{Labbe2006,
  Labbe2010, Labbe2015}, the deblended \emph{Spitzer}/IRAC photometry was
measured with \textsc{mophongo}, using the \emph{HST} observations as a
template, in the deep $\sim 200$-hour data from the GREATS program (M.~Stefanon
et al., subm.).  The latter provides constraints on the rest-frame optical part
of the spectral energy distribution redward of the 4000-\AA\ break and is
critical to pin down the stellar masses of our galaxies.  The results are
listed in \autoref{tab:sources}, for the galaxies in the systemic redshift
sample.

As part of the DR2, the spectra of all galaxies are modeled with
\textsc{pyplatefit} (R. Bacon et al., in prep.), the Python
implementation of the spectrum fitting code \textsc{platefit}, originally
developed for the \emph{Sloan Digital Sky Survey} \citep{Tremonti2004,
  Brinchmann2004, Brinchmann2008}.  The most salient features of
\textsc{pyplatefit}, relevant to this work, are that it can simultaneously
model both the emission- and absorption lines, as well as the stellar
continuum, allowing for velocity differences between groups of lines (such as
\Lya\ and other UV lines).  All lines are modeled using Gaussians except \Lya,
for which a (double) asymmetric Gaussian\footnote{Also known as the Skew normal
  distribution, $f(x) = 2 \phi(x)\Phi(\gamma x)$, where $\phi(x)$ is the
  standard normal (Gaussian) distribution, $\Phi(x)$ is the cumulative
  distribution function for a standard normal distribution, and $\gamma$ is the
  skewness parameter.}  is used (double if the \Lya\ line is double-peaked).

\subsection{Measurement of systemic redshifts}
\label{sec:meas-syst-redsh}
We obtain systemic redshifts for galaxies in our parent sample from either the
rest-frame UV features using MUSE (\autoref{sec:muse}) or the rest-frame
optical emission lines with near-IR spectroscopy (\autoref{sec:kmos} and
\autoref{sec:mosfire}).  For the near-IR follow-up, targets were selected by
their \Hmag\ magnitude (as a proxy for stellar mass) and the absence of a
systemic redshift from MUSE in the DR1 reductions.  Fainter targets were
sometimes observed because brighter targets were no longer accessible given the
small size of the HUDF and physical limitations in the positioning arms and
slits of multi-object spectrographs.
\begin{figure*}[t]
  \centering
  \includegraphics[width=1.0\textwidth]{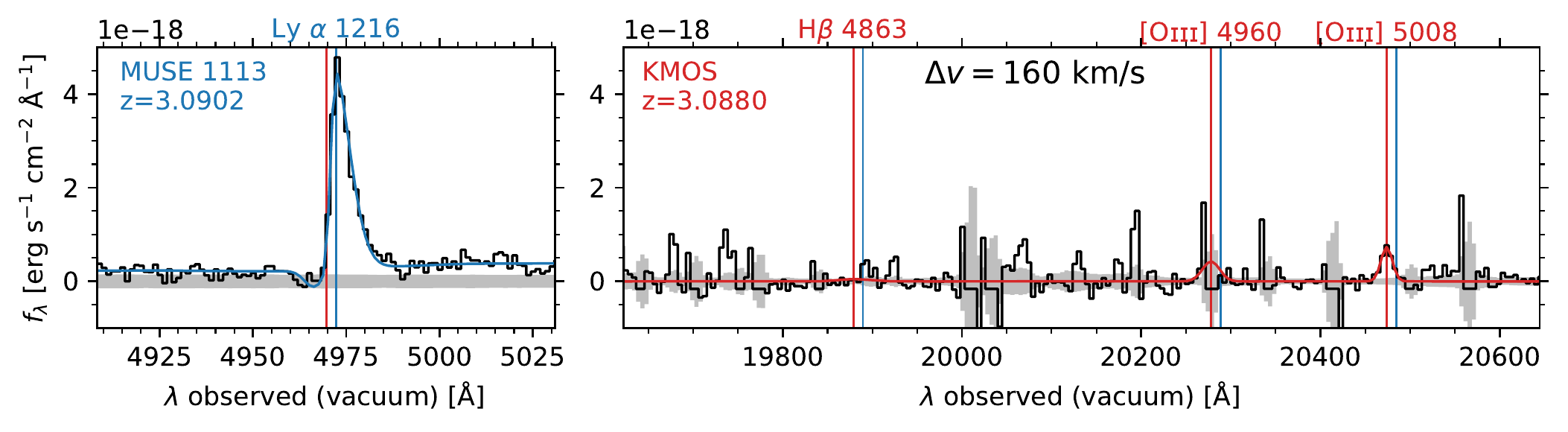}
  \caption{Rest-frame ultraviolet and optical spectra for the first galaxy in
    the sample with near-infrared follow-up.  The \textbf{left} panel shows the
    MUSE spectrum surrounding the \Lya\ line.  The \textbf{right} panel shows
    the continuum subtracted KMOS spectrum around the \Hbeta\ and \Oiii\ lines.
    In both panels the vertical blue and red lines indicate the redshift of
    \Lya\ and the systemic redshift, respectively, determined from the fit to
    the spectrum (shown in the same color).  This particular galaxy is detected
    in \Lyalpha\ and \Oiiib, but not in \Hbeta, showing a positive velocity
    offset between the red peak of \Lya\ and the systemic redshift (note
    \Oiiia\ falls on top of a skyline).  Spectra of the remaining galaxies with
    KMOS or MOSFIRE observations are shown in \autoref{fig:spectra} in
    \autoref{sec:spectra}.
    \label{fig:spectrum_example}}
\end{figure*}
\subsubsection{MUSE}
\label{sec:muse}
For a subset of galaxies we can determine the systemic redshift directly from
the MUSE spectra, using the weaker rest-frame UV emission lines, or absorption
features.  We identify objects in the parent sample that are cataloged as
having $\mathrm{S/N}>3$ in at least one UV emission line.  We focus
specifically on \Oiiiuv\, \Siiii, \Ciii\ and a selection of absorption
lines,\footnote{\Siiia, \Oi, \Siiib, \Cii, \Siiv, \Feii, \Alii, and \Aliii.}
that trace the systemic redshift.  We also fit narrow \Heii\ together with the
emission lines, finding it at a similar velocity offset as the other UV lines.
We do not use the resonant lines, such as \Civ, which can be offset from the
systemic velocity like \Lya.  To identify absorption line redshifts, we inspect
all objects with $\Vmag \le 27$ and/or $\Imag \le 27$, finding that we can
determine these in several galaxies down to $\Imag = 26$.  We use
\textsc{pyplatefit} to fit the selected spectra, performing 200 bootstrap
iterations to obtain a more robust estimate of the uncertainties (both on \Lya\
and the other features).  We only keep the objects that remain at
$\mathrm{S/N} > 3.5$ in at least one emission line or the sum of the absorption
features.  In addition, we keep MUSE-1360 as a tentative candidate, having both
a tentative detection in the KMOS data and an absorption line redshift at
$\mathrm{S/N} = 3$.  The \zLya\ and \zsys, with their bootstrapped
uncertainties, are provided in \autoref{tab:sources} (where $\zLya$ is the
redshift of the red peak of \Lya).  We note that three of these sources were
also part of the study of \Lya-velocity offsets by \cite{Verhamme2018}.

In principle, one could estimate the systemic redshift by using half of the
peak separation for \Lya-lines with a blue bump \citep{Verhamme2018}.  Indeed,
a few galaxies in our sample also show blue bump-emission.  However, systematic
searches for blue-bump \Lya-emitters are still on-going and we therefore do not
include such a sample at this stage.  Furthermore, the presence of a blue bump
requires specific radiative transfer conditions and selecting a sample in such
a way may introduce a bias in the stack.

\subsubsection{KMOS}
\label{sec:kmos}
\begin{figure*}[t]
  \centering
  \includegraphics[width=\textwidth]{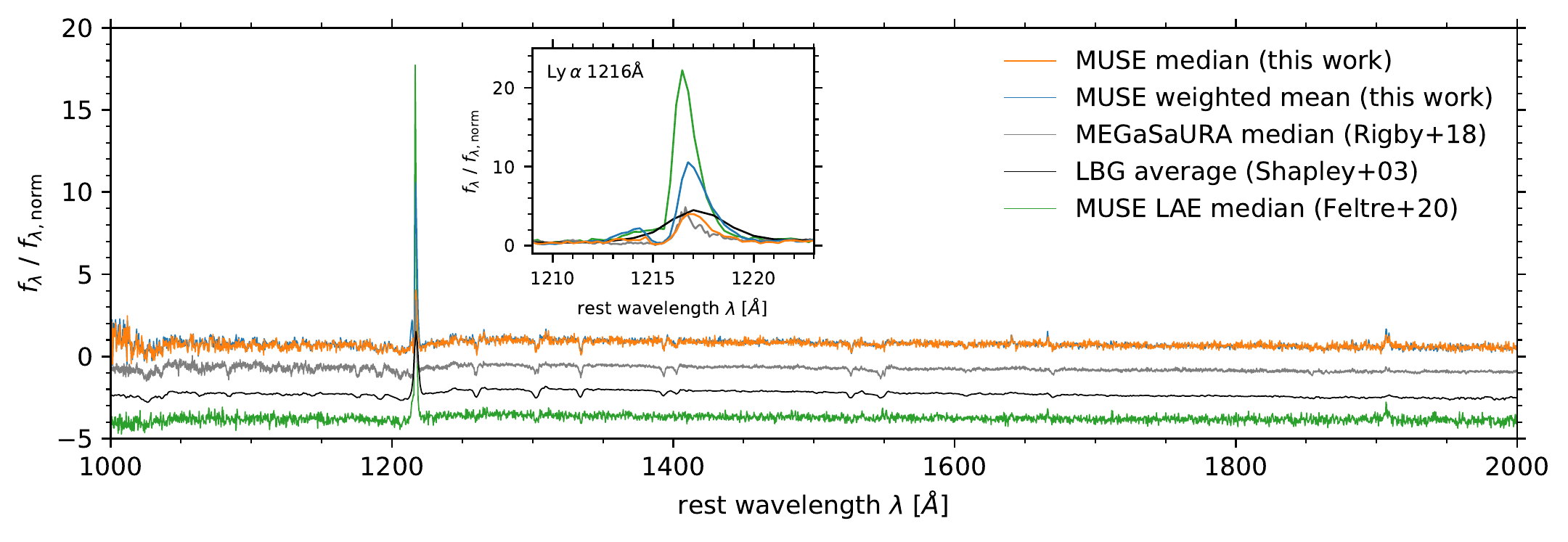}
  \caption{Composite MUSE spectrum of the sample, from both weighted mean
    (blue) and median (orange) stacking (at 0.3\,\AA\ resolution) using the
    systemic (non-\Lya) redshifts.  We compare to the average Lyman Break
    Galaxy (LBG) spectrum at $z\sim3$ (black) from \cite[][1\,\AA\
    resolution]{Shapley2003}, the composite spectrum of 14 strongly lensed,
    star-forming galaxies at $1.6 < z < 3.6$ from the
    \textsc{Meg}aSa\textsc{ura} sample (\citealt{Rigby2018b}; the pivot
    normalized, median stack at 0.1\,\AA\ resolution) and the composite MUSE
    spectrum of all 220 LAEs at $2.9 < z < 4.6$ \citep[][median stack; note
    their weighted mean stack is very similar, with slightly stronger \Lya\
    emission]{Feltre2020}.  We normalize all spectra to the median flux density
    at 1267--1276\,\AA\ and offset the literature spectra by $-1.5, -3$ and
    $-4.5$ for clarity.  The inset shows a zoom-in of the \Lya-line without
    vertical offsets. Overall, the composite MUSE spectrum is very similar to
    the LBGs and \textsc{Meg}aSa\textsc{ura}, showing comparable UV continuum
    and absorption features, slightly stronger UV emission lines, and a
    comparable strength of the \Lya-line in the median spectrum (notably
    showing a blue bump).  In contrast, the median stack of all LAEs from
    \cite{Feltre2020} shows significantly stronger \Lya-emission, even when
    compared to the weighted mean spectrum of our
    galaxies.\label{fig:muse_stack}}
\end{figure*}
The KMOS observations were taken in two ESO periods, as part of 099.A-0858(A)
(PI: Bouwens) and 0101.A-0725(A) (PI: Boogaard).  We used the HK grating (with
a spectral resolution of $\lambda/\Delta \lambda \approx 1800$) in 5 Observing
Blocks (OBs) per period, with an ABA ABA AB sky-offset pattern between the
science (A) and sky (B) frames with 300\,s integrations and 0\farcs2 dithering
offsets.  In total we targeted 17 galaxies, with (final) on source times
ranging between 200 and 250 minutes.  We also included a bright quasar on all
masks to control the astrometry, from which we measure the image quality to be
around 0\farcs75 and 0\farcs85 in the reduced P99 and P101 data, respectively.

We reduce the data using the ESO KMOS pipeline version \texttt{2.1.0}
\citep{Davies2013}, using the standard star observations for the zeropoint,
response and telluric correction.  We enable the \texttt{background} flag to
correct for differences in the residual background level between the exposures
by applying a constant offset, estimated by taking the mode of the pixel values
after excluding the brightest 25\%.  We discard the data from one detector for
the second A frame of the first OB in period 99, which shows a strong
background offset.  We experimented with further reducing the sky line
residuals using the \texttt{sky-tweak} and \texttt{molecfit} options of the
pipeline, but found that these sometimes introduces artifacts in the data.  As
our lines were selected to be away from the sky lines as much as possible, we
therefore do not apply these corrections.  Because of the density of the
skylines around \Oiialt\ in the H-band, we focus on the \Hbeta\ and \Oiii\ in
the K-band.  We apply radial velocity
corrections\footnote{$\lambda' = \lambda \sqrt{(1 + v_{r}/c)/(1 - v_{r}/c)}$,
  where $\lambda'$ and $\lambda$ are the corrected and uncorrected wavelengths,
  respectively, $c$ is the speed of light and $v_{r}$ is the radial velocity
  correction to the solar system barycenter, computed with
  \texttt{astropy.coordinates.SkyCoord.radial\_velocity\_correction}.\label{ft:bary}}
to shift every reduced A-B frame to the solar system barycentric frame (the
mean $\avg{v_{r}} = 17.3$~km~s$^{-1}$).

We correct for positional shifts between the different OBs by centering on the
continuum position of the reference quasar, which we model with a 2D Gaussian.
As objects were placed on different IFUs between periods, their position
relative to the reference quasar change.  We therefore first combine and
analyze the data from each period separately.  To identify the spatial position
of each (emission line-only) object on its IFU in each period, we
\begin{inlinelist}
\item extract spectra at the a-priori expected position (that is, the quasar
  position or the center of the cube) using the 2D fit of the reference quasar
  as a spatial model,
\item identify the brightest spectral line, \Oiiib, based on the \Lya\ redshift
  and determine its central wavelength and line width using a Gaussian fit,
\item collapse the cube over the channels with line emission to make a `narrow
  band', through multiplication with the Gaussian fit along the wavelength
  axis, and
\item identify the spatial position in the narrow band image.
\end{inlinelist}
We iterate steps (i)-(iv) until we converge on spatial position.  Finally, we
combine the data from both periods using the best positions and repeat the same
steps to obtain the final spectra.

We conservatively only consider the objects for which we can identify the
line(s) in each half of the data separately, which gives strong confidence that
the line(s) are not (caused by) sky line residuals.  We exclude one source
where the blueshift of the lines relative to \Lya\ resulted in them being too
close to the skylines to determine the centroid and four more sources where a
tentative feature was only seen in one period.  In total, we confidently detect
the rest-frame optical line(s) in 7/17 galaxies.  As an example, we show the
MUSE and KMOS spectrum for one of the galaxies in
\autoref{fig:spectrum_example}.  The spectra of the remaining galaxies are
shown in \autoref{fig:spectra} in \autoref{sec:spectra}.

Finally, we determine the redshift by simultaneously fitting Gaussians (in
vacuo) to the \Hbeta\ and \Oiii\ lines (using \textsc{lmfit};
\citealt{lmfit_1_0_0}).  We use the inverse of the error spectrum as weights
and subtract a running median continuum from the spectrum prior to the fitting.
The resulting redshifts are reported in \autoref{tab:sources}.
\subsubsection{MOSFIRE}
\label{sec:mosfire}
The MOSFIRE observations were taken in the night of 28 November 2018 as part of
2018B\_N182 (PI: Riechers).  We observed a single K band mask with 0\farcs7
slits ($\lambda / \Delta \lambda \approx 3610$).  We used an AB dither pattern
with 180 second exposures, totaling to 108 minutes of exposure time on source,
with an average seeing of 0\farcs7.  The data were reduced using the standard
MOSFIRE Data Reduction Pipeline (Release
2018),\footnote{\url{https://github.com/Keck-DataReductionPipelines/MosfireDRP}}
using the Neon arc lamps for the wavelength calibration.  As our objects
generally do not show any continuum, we first manually identify (candidate)
emission lines in the rectified, two dimensional spectra (based on the \Lya\
redshift).  We then optimally extract the one dimensional spectra using a
Gaussian model for the spatial profile.  As all data were taken on a single
night, we apply the radial velocity correction to the final spectra
($v_{r} = -9.2$\,km\,s$^{-1}$).\footnote{See footnote \ref{ft:bary}}

In total, we detect the rest-frame optical lines in 4/7 of the galaxies on the
mask that are part of our parent sample (including MUSE-6518, with blended
\emph{HST} photometry).  Their spectra are shown in \autoref{fig:spectra}.  We
measure the redshifts as described in \autoref{sec:kmos} and report the results
in \autoref{tab:sources}.
\subsection{Final systemic redshift sample}
\label{sec:final-sample}
In total, we use MUSE, KMOS and MOSFIRE to obtain systemic redshifts for \Nobj\
galaxies, \Nlae\ of which are originally identified by their \Lya-emission,
with an average redshift of $\avg{z} = \zavg$.  The \Hmag\ magnitude of the
final sample is shown in comparison to the MUSE parent sample in
\autoref{fig:z-f160}.  We have a systemic redshift for most galaxies in the
parent sample down to $\Hmag =26$.  Because the parent sample is \Lya-selected,
this raises the question how representative our sample is for the broader
population of galaxies at these epochs.  We therefore compare our (parent)
sample to all galaxies at the same redshift and over the same field, from the
photometric redshift catalog by \cite[][updated with the MUSE
redshifts]{Rafelski2015}, after excluding the X-ray AGN \citep[][as we did for
the parent sample]{Luo2017}, see \autoref{fig:z-f160}.

We show a histogram of the physical properties of the galaxies in
\autoref{fig:hist}.  The median stellar mass and SFR of the sample is $\Mmed$
and $\SFRmed$.  The sample encompasses $\geq 50\%$ of the galaxies in the MUSE
parent sample in the bins down to $M_{*} \geq 10^{9}$\,\Msun\ and
$\mathrm{SFR} \geq 3$\,\Msun\,yr$^{-1}$, and $\geq 20\%$ of the galaxies in the
broader photometric catalog, down to the same limits.

Our galaxies are faint \Lya-emitters in comparison to narrow band-selected
samples.  The typical \Lya\ luminosity of our sample is
$L_{\Lya} = \LLyaavg \approx 0.2\,L_{\Lya}^{*}$ at $z=3.5$
\citep[e.g.,][]{Ouchi2008, Drake2017, Herenz2019}.  The average rest-frame
equivalent-width\footnote{The rest-frame equivalent width is computed by
  \textsc{pyplatefit}, from the total flux in \Lya\ (including a possible blue
  bump) over the modeled continuum flux density at 1216\AA\ (defined such that
  a positive value indicates emission).} of our sample,
$\avg{\EWLya} \approx \EWavg$\,\AA, is also small, especially when considering
that $\approx 25$\,\AA\ is the typical lower limit for the definition of a
(narrow band-selected) \Lya-emitter (LAE; e.g., \citealt{Matthee2016b}).  This
is likely due to our selection towards objects that are bright in \Hmag.  As
such, our galaxies are not necessarily comparable to the typical sample of
LAEs, but arguably more similar to the average population of (low-mass)
star-forming galaxies

To illustrate this point, we stack the MUSE spectrum of all the galaxies in our
sample, using the systemic redshifts, following \cite{Feltre2020}.  We perform
a median and weighted mean stack, after normalizing each of the spectra by the
median flux density at $1267 - 1276$\,\AA\ (matching \citealt{Rigby2018b}).
The result is shown in \autoref{fig:muse_stack}.  We compare the composite MUSE
spectrum to the average spectrum of $z\sim3$ Lyman Break Galaxies \citep[LBGs;
][]{Shapley2003}, the composite spectrum of 14 strongly lensed, star-forming
galaxies at $1.6 < z < 3.6$ from the \textsc{Meg}aSa\textsc{ura} sample
\citep{Rigby2018b} and the composite MUSE spectrum of all 220 LAEs at
$2.9 < z < 4.6$ from \cite{Feltre2020}.  The median spectrum of the galaxies in
our sample shows significantly weaker \Lya-emission than the median spectrum of
all MUSE LAEs (that includes a majority of LAEs with a fainter $\Hmag > 26$).
Instead, the similarities in the literature spectra of LBGs and star-forming
galaxies and the (median) composite spectrum of our sample support the case
that our galaxies are more similar to the population of star-forming galaxies
at these epochs.
\section{Results}
\label{sec:results}
\begin{figure}[t]
  \includegraphics[width=\columnwidth]{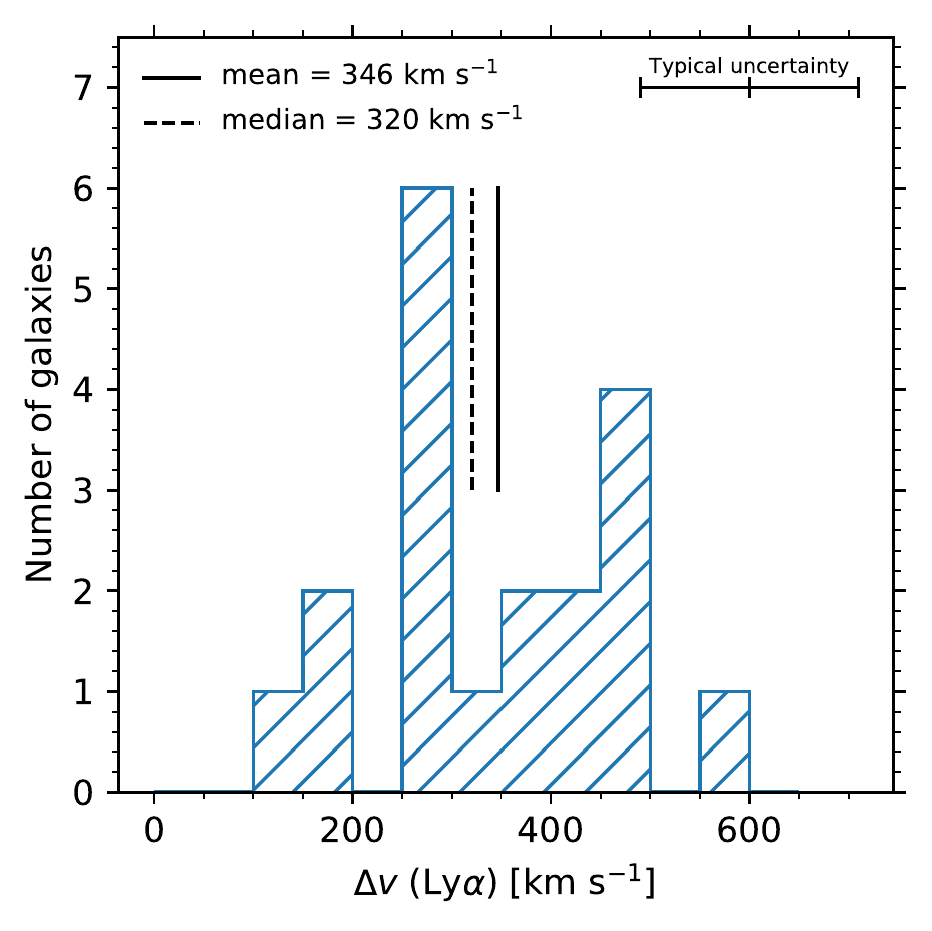}
  \caption{Histogram of the velocity offset of \Lyalpha\ with respect to the
    systemic redshift, $\Delta v(\Lya)$.  The vertical lines indicate the mean
    and median velocity offset in the sample.  The typical
    $\pm 1\sigma$-uncertainty on the velocity offset is indicated in the top
    right.\label{fig:vel_offs}}
\end{figure}
\begin{figure*}[t]
  \centering
  \includegraphics[width=0.9\textwidth]{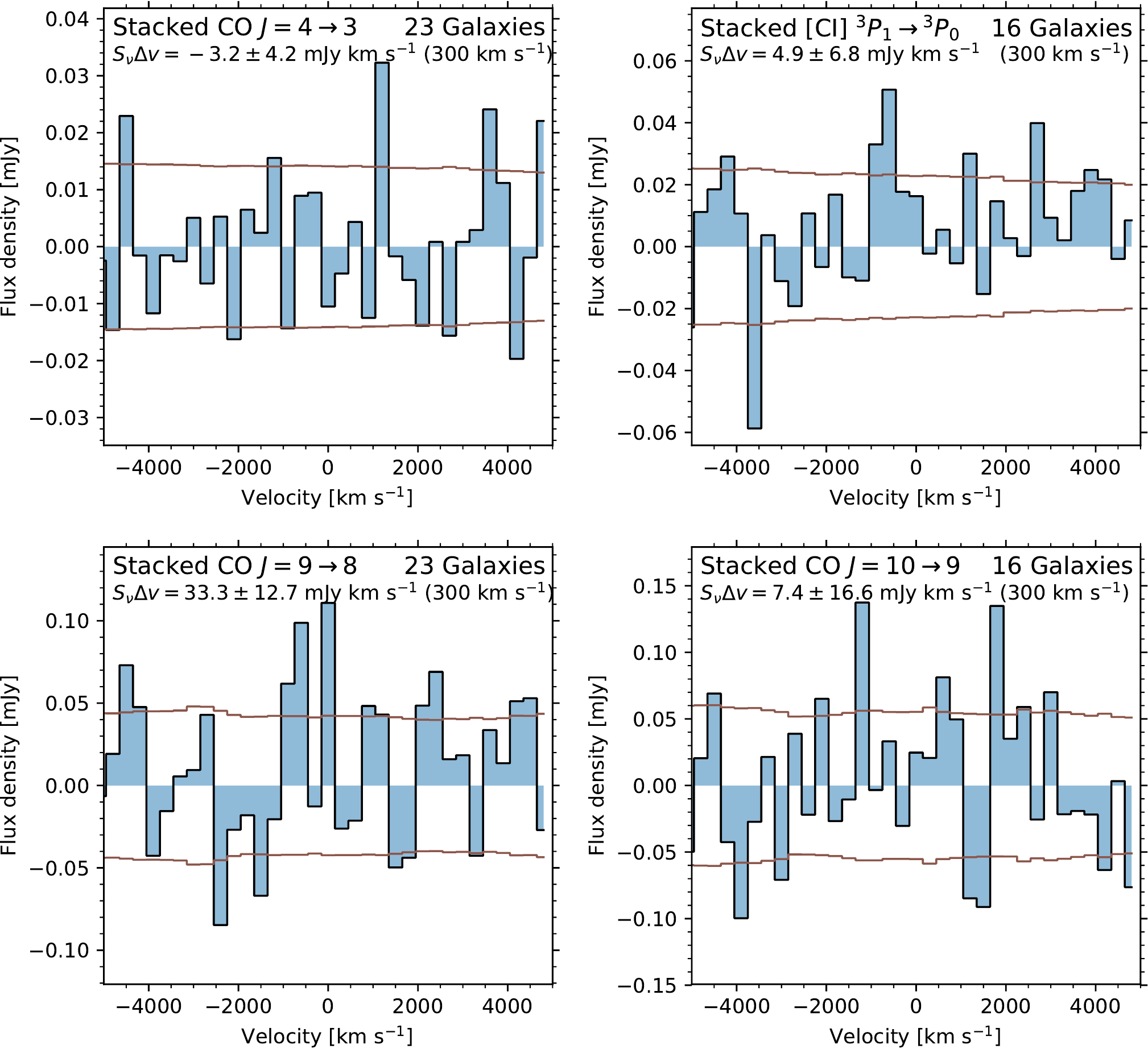}
  \caption{Stacked spectra of the $^{12}$CO and \CI\ transitions of the full
  sample.  The spectrum is shown in blue and the brown line shows the
  root-mean-square noise (propagated through the $1/\sigma^{2}$-weighting of
  the stack).  The stacked transition, the number of galaxies in the stack, the
  integrated line flux in the zero-velocity channel and the channel width
  (\deltavstack~km~s$^{-1}$) are indicated in each panel.  We do not detect any
  emission in any of the transitions and provide $3\sigma$ upper limits in
  \autoref{tab:lines}. \label{fig:stacks_all}}
\end{figure*}
\subsection{Velocity offsets}
\label{sec:velocity-offsets}
We plot the velocity offset of \Lyalpha\ with respect to the systemic redshift
in \autoref{fig:vel_offs}, defined as
$\Delta v(\Lya) = c (\zLya - \zsys)/(1+\zsys)$, where $c$ is the speed of
light.  Our galaxies show a mean velocity offset of
$\avg{\Delta v(\Lya)} = \deltavlya$\,km\,s$^{-1}$, with a range from 100 to 600
km\,s$^{-1}$.

For comparison, the mean $\Delta v(\Lya) \approx 200 - 250$\,km\,s$^{-1}$ in
the LAE samples at $z=2-3$ \citep{Erb2014,Trainor2015}, while the $z\sim3$ LBG
sample from \cite{Shapley2003} shows a greater mean velocity offset of
650\,km\,s$^{-1}$.  The relatively large velocity offsets imply a larger \HI\
column density and lower \Lya\ escape fraction, consistent with the low \EWLya\
of our galaxies \citep{Shapley2003, Erb2014, Yang2017a}.

The broad distribution in \autoref{fig:vel_offs} also reflects the smoothing
function by which the stacking signal would be diluted if $\zLya$ would be used
for stacking, in particular because the ALMA data has a higher velocity
resolution.  This highlights the need for systemic redshifts.
\subsection{ALMA Stacking}
\label{sec:stacking}
\begin{deluxetable*}{lccccccc}
  \tablecaption{Stacking results for the CO and \CI\ lines. \label{tab:lines}}
  \tablehead{
    \colhead{Transition} &
    \colhead{$\nu_{0}$} &
    \colhead{$z_{\mathrm{min}}$} &
    \colhead{Band} &
    \colhead{N} &
    \colhead{$S_{\nu}\Delta v$} &
    \colhead{$S_{\nu} \Delta v$} &
    \colhead{$L'$} \\
    \colhead{} &
    \colhead{(GHz)} &
    \colhead{} &
    \colhead{} &
    \colhead{} &
    \colhead{(mJy~km~s$^{-1}$)} &
    \colhead{(mJy~km~s$^{-1}$)} &
    \colhead{(K~km~s$^{-1}$~pc$^{2}$)}
  }
  \colnumbers
  \startdata
 \multicolumn{8}{c}{Systemic redshift sample (\autoref{fig:stacks_all})}\\
 \hline
 CO $J=4 \rightarrow 3$                     & 461.04  & 3.0115 &    3 & \cofourstacknum & \cofourstackflux & $<\cofourstacklimit$ & $<\cofourstacklimitLp$\\
 \CIthreePone & 492.16  & 3.2823 &    3 &  \cistacknum    & \cistackflux     & $<\cistacklimit$     & $<\cistacklimitLp$      \\
 CO $J=9 \rightarrow 8$                     & 1036.91 & 2.8122 &    6 & \coninestacknum & \coninestackflux & $<\coninestacklimit$ & $<\coninestacklimitLp$\\
 CO $J=10 \rightarrow 9$                    & 1151.99 & 3.2352 &    6 & \cotenstacknum & \cotenstackflux  & $<\cotenstacklimit$  & $<\cotenstacklimitLp$\\
\hline
 \multicolumn{8}{c}{\Lya-selected galaxies only}\\
 \hline
 CO $J=4 \rightarrow 3$                     & 461.04  & 3.0115 &    3 & \cofourstacknumlae & \cofourstackfluxlae & $<\cofourstacklimitlae$ & $<\cofourstacklimitLplae$\\
 \CIthreePone & 492.16  & 3.2823 &    3 &  \cistacknumlae    & \cistackfluxlae     & $<\cistacklimitlae$     & $<\cistacklimitLplae$      \\
 CO $J=9 \rightarrow 8$                     & 1036.91 & 2.8122 &    6 & \coninestacknumlae & \coninestackfluxlae & $<\coninestacklimitlae$ & $<\coninestacklimitLplae$\\
 CO $J=10 \rightarrow 9$                    & 1151.99 & 3.2352 &    6 &  \cotenstacknumlae & \cotenstackfluxlae  & $<\cotenstacklimitlae$  & $<\cotenstacklimitLplae$
 \enddata
\tablecomments{$\avg{z}=\zavg$, $\Delta v = \deltavstack$~km\,s$^{-1}$. (1)
  Stacked transition (2) Rest frequency (3) Minimum redshift at which
  the transition is covered by ASPECS. (4) Band that contains transition. (5)
  Number of objects in stack. (6) Line flux in stack. (7) $3\sigma$ upper limit on
  line flux. (8) $3\sigma$ upper limit on line luminosity.}
\end{deluxetable*}
With the systemic redshifts in hand, we turn to the ALMA data.  We use the
ASPECS Band 3 \citep{Gonzalez-Lopez2019, Decarli2019} and Band 6
\citep{Gonzalez-Lopez2020, Decarli2020} datacubes at their native resolution
($\approx 20$\,km\,s$^{-1}$ in both cases).  The root-mean-square (rms) error
spectra reach $\approx 0.2$ and 0.5\,mJy\,beam$^{-1}$\,channel$^{-1}$ in Band 3
and Band 6, respectively (at the center of the field, varying with frequency).

Before extracting the spectra, we first shift the ALMA cubes from the Kinematic
Local Standard of Rest (LSRK) to the Barycentric frame, using the \textsc{CASA}
task \texttt{imreframe} ($\Delta v = -16.78$\,km\,s$^{-1}$), such that all our
spectroscopic data are on the same velocity frame.  We then extract pixel
spectra at the \emph{HST} positions \citep{Rafelski2015} of our galaxies, after
correcting for the known astrometric offset
($\Delta \alpha = 0\farcs076, \Delta \delta = - 0\farcs279$;
\citealt{Dunlop2017}, consistent with \citealt{Franco2020b}).  These spectra
should contain all the flux as our sources are expected to be unresolved by the
ASPECS synthesised beam ($1\farcs8 \times 1\farcs5$ in Band 3 and
$1\farcs5 \times 1\farcs1$ in Band 6).  Their spatial extent in the rest-frame
UV is significantly smaller, with a median effective radius in \Hmag\ of
$\approx 0\farcs2$ \citep{vanderWel2012}.

Inspecting the spectra around the systemic redshift, none of the galaxies are
individually detected in their CO or \CI\ emission lines at the $3 \sigma$
level.  We therefore stack the spectra as follows \cite[][cf.
\citealt{Spilker2014}]{Boogaard2020}.  We first create a grid in velocity
space, centered around zero, with \deltavstack~km~s$^{-1}$ wide channels.  The
channel width was chosen based on the mean rest-frame UV/optical line-width
($\approx 200$~km~s$^{-1}$), such that $> 90\%$ of the stacked line flux is
expected to fall within the single central channel.  We convert each observed
spectrum to velocity space, centered around the line, and bin it onto the
velocity grid.  We then stack the spectra by taking the $1/\sigma^{2}$-weighted
mean in each velocity channel (where $\sigma$ is the error) and determine the
uncertainty by propagating the error spectrum in the same manner.  We finally
measure the flux density and the corresponding uncertainty in the zero-velocity
channel.

The stacked spectra are shown in \autoref{fig:stacks_all}.  None of the lines
are detected in the stack at a signal-to-noise ratio greater than three.  The
stack of CO $J=9\rightarrow8$ shows some signal at the $2-3\sigma$ level, but
we do not consider this a detection.  The high-$J$ lines of CO are primarily
sensitive to the gas heating (and not the gas mass) and we include the
constraints on these lines mainly for completeness and future reference.  We
compute $3\sigma$ upper limits on the integrated line flux, $S_{\nu} \Delta v$,
from the uncertainty in the zero-velocity channel of the stacked spectrum (at
300\,km\,s$^{-1}$ resolution).  We determine the corresponding upper limits on
the line luminosities via
\begin{align} L' = 3.255 \times 10^{7}\, {S_{\nu}\Delta v}\, d_{L}^{2}
\nu_{\mathrm{obs}}^{-2} (1+z)^{-3} \,
\mathrm{K\,km\,s}^{-1}\,\mathrm{pc}^{2} \label{Llineprime}
\end{align}
\citep{Solomon1992, Carilli2013}, adopting the luminosity distance ($d_{L}$)
and observed frequency ($\nu_{\mathrm{obs}}$) at the average redshift of the
sample, $z=3.45$.  The results can be found in \autoref{tab:lines}.  We also
perform additional stacks, including only the galaxies with the highest stellar
masses ($M_{*} \geq 10^{9.0}$, $\geq 10^{9.5}$ and $\geq 10^{10}$\,\Msun) and
the highest star formation rates ($\mathrm{SFR} \ge 3.2$ and
$\geq 10$\,\Msun\,yr$^{-1}$), but do not obtain any detections.  Stacking all
the different CO lines together does not yield a detection either (regardless
of whether \CI\ was also added to this stack).

\begin{figure}[t]
  \centering
  \includegraphics[width=0.9\columnwidth]{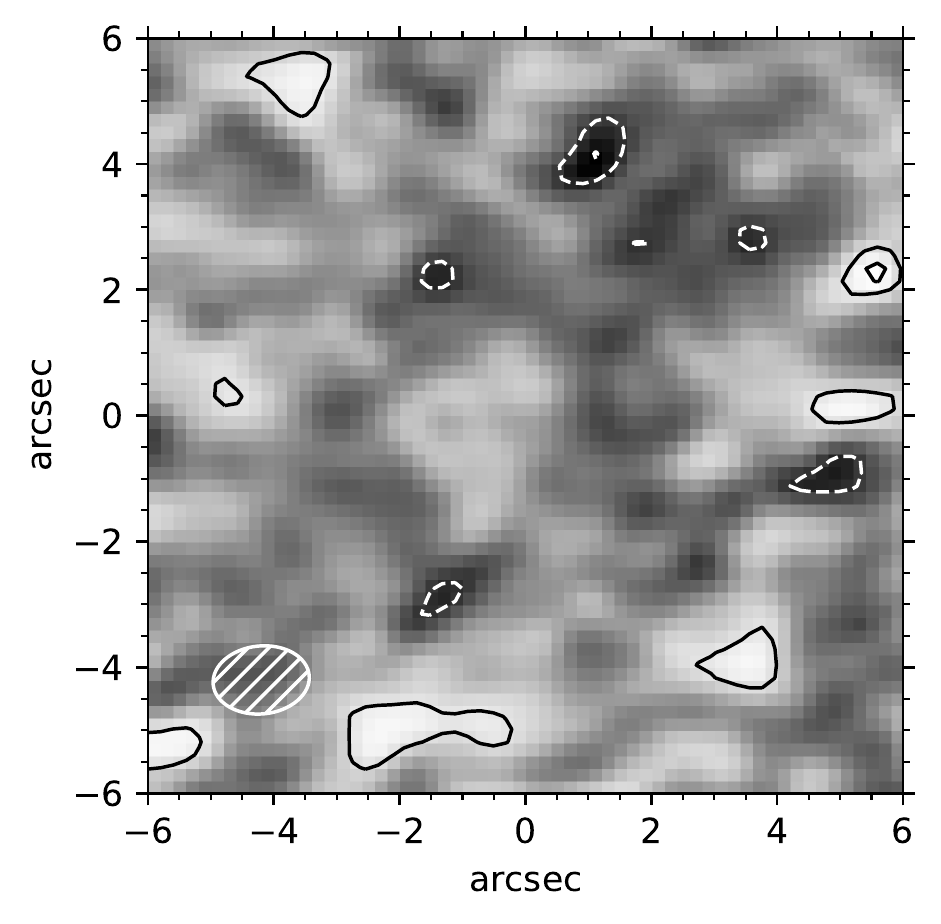}
  \caption{Stack of the 1.2\,mm dust continuum.  The cutout size is
    $12'' \times 12''$ and the synthesised beam is indicated in the bottom left
    corner.  Contours are drawn starting at $\pm 2\sigma$ in steps of
    $\pm 1\sigma$ (dashed contours indicate negative signal).  No emission is
    detected at the $3\sigma$ level, implying an upper limit of
    $\dustlimit$. \label{fig:duststack}}
\end{figure}

None of the galaxies are individually detected at the $3 \sigma$ level in the
deep 1.2\,mm dust continuum map \citep{Gonzalez-Lopez2020, Aravena2020}.  In
addition to the line stack, we therefore also perform a weighted mean stack of
the 1.2\,mm dust continuum data for the full sample \citep[following the
approach from][again applying the astrometric offset]{Bouwens2016,
  Bouwens2020}.  We do not obtain a detection, measuring a flux of
$S_{\nu} = \duststack$ (\autoref{fig:duststack}), implying an upper limit of
$\leq \dustlimit$ ($3 \sigma$).  Following the same procedure for the 3\,mm
continuum results in an upper limit of $\leq \dustlimitThreemm$ ($3\sigma$).
In the following, we focus on the limit from the deep 1.2\,mm continuum, as it
provides the strongest constraints on the mass in dust and gas.
\section{Discussion}
\label{sec:disc}
\subsection{Molecular gas masses}
\label{sec:molecular-gas-masses}
\begin{figure*}[t]
  \centering
  \includegraphics[width=0.75\textwidth]{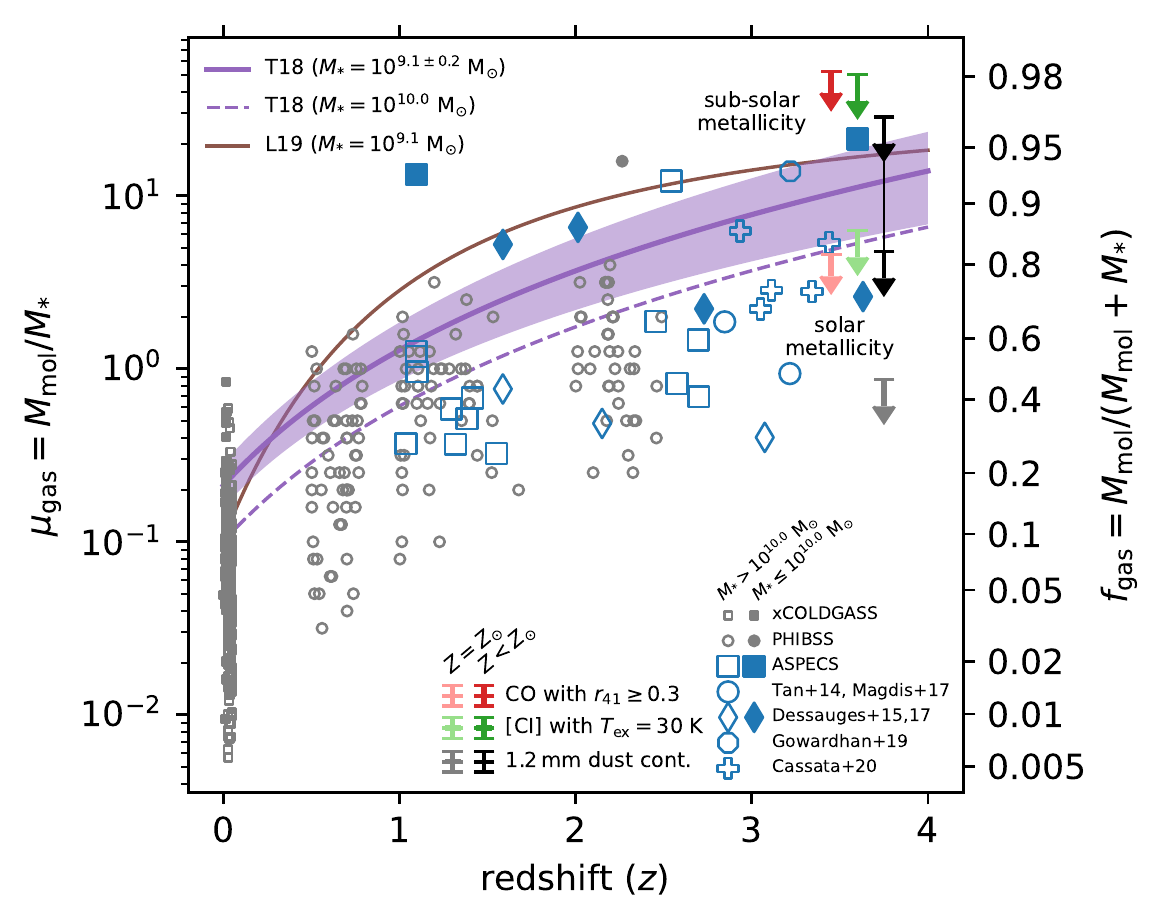}
  \caption{The molecular gas-to-stellar mass ratio (\mugas) and gas fraction
    (\fgas) as a function of redshift.  We show a literature sample of
    star-forming galaxies with CO measurements, focusing on $z\ge3$, separating
    galaxies with $M_{*} \le 10^{10}$\,\Msun\ (filled symbols) and
    $M_{*} > 10^{10}$\,\Msun (open symbols).  Nearly all galaxies from
    literature are significantly more massive than the $\avg{z} = \zavg$ star
    forming galaxies studied here, which have a median $M_{*} = \Mmed$ and
    $\mathrm{SFR} = \SFRmed$.  The upper limit on the gas fraction from CO is
    shown in red, with the limits at low and high gas fraction corresponding to
    solar metallicity ($\aco^{\mathrm{MW}} = 4.36$; light red) and sub-solar
    metallicity ($\aco = \acoavg$; dark red), respectively, assuming
    $r_{41} \ge 0.3$.  The green limit shows the constraints from \CI, for a
    typical abundance of $2\times10^{-5}$ and a factor $8\times$ lower in
    lighter and darker shading, respectively.  The gray and black limits show
    the constraint from the 1.2\,mm dust continuum assuming, in order of
    increasing gas fraction, $\gdr=100, \gdravglin$ and $\gdravgbpl$,
    corresponding to solar metallicity and sub-solar metallicity with different
    assumptions for the scaling of $\gdr \propto Z^{\gamma}$.  We add
    horizontal offsets to the upper limits from different tracers for clarity.
    For comparison, we also show the predictions from \citet[][based on CO and
    dust-continuum measurements]{Tacconi2018} for main-sequence galaxies with
    stellar masses $M_{*} = 10^{9.1 \pm 0.2}$\,\Msun\ (solid line with shading;
    0.2 dex around the median mass of our sample, where shading also includes
    the uncertainties in the fit of the scaling relation) and
    $10^{10.0}$\,\Msun\ (dashed; above the most massive galaxy in our sample),
    and from \citet[][based on dust continuum only]{Liu2019b} for
    $M_{*} = 10^{9.1}$\,\Msun\ (solid brown line).  Under the assumption of
    solar metallicity conversion factors, the constraints are in tension with
    the scaling relations for galaxies in our mass range, while for sub-solar
    metallicity conversion factors the upper limits are in comfortable
    agreement.  \label{fig:fgas}}
\end{figure*}
\begin{figure}[t]
  \includegraphics[width=\columnwidth]{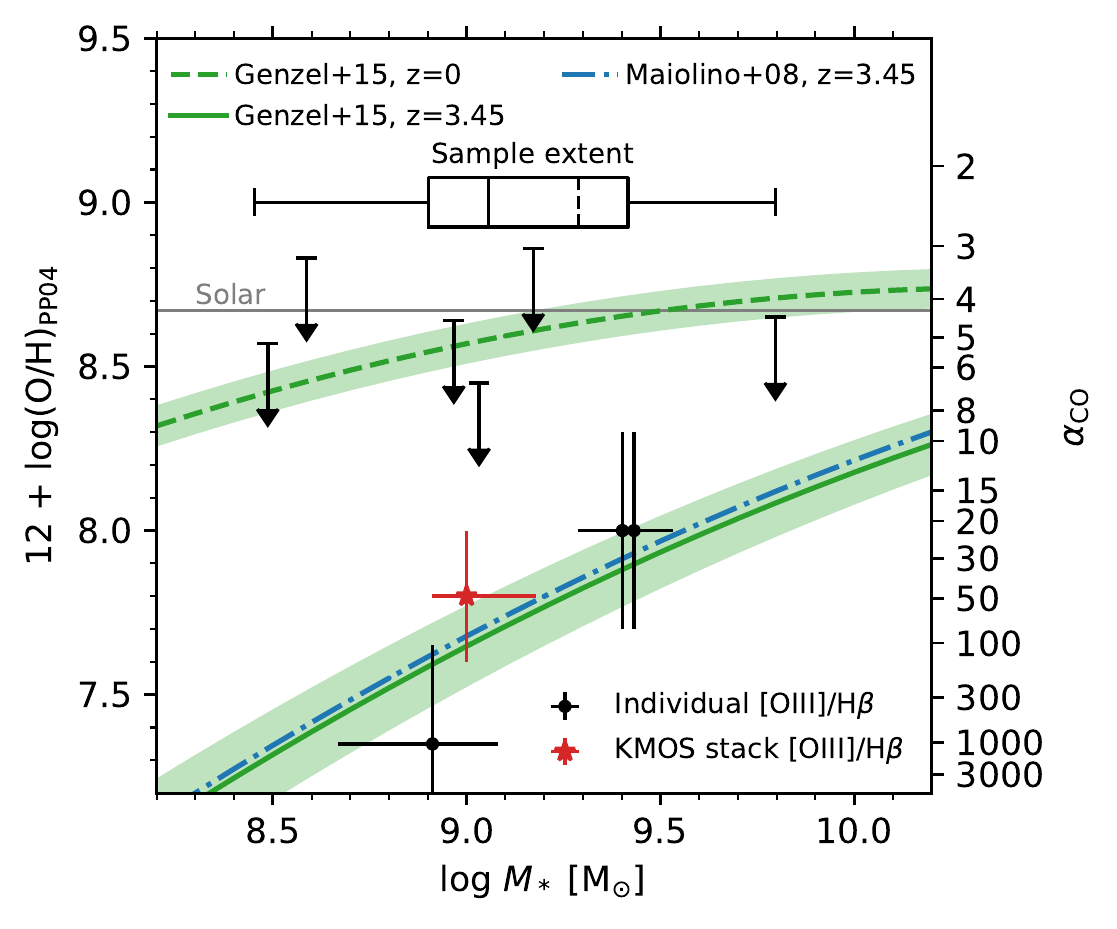}
  \caption{Plot of the mass--metallicity (MZ) relation and corresponding
    $\aco(Z)$ (Equations \ref{eq:aco} and \ref{eq:MZ}; \citealt{Genzel2015}),
    as function of stellar mass ($M_{*}$).  The constraints on the MZ relation
    at $z=3.45$ come from \cite{Maiolino2008}.  Metallicity is plotted as
    \logOH\ on the \cite{Pettini2004} scale.  The inset shows a boxplot with
    the whiskers indicating the full stellar mass extent of the sample, the box
    the interquartile range, the solid line the median, and the dashed line the
    (linear) mean $M_{*}$ of the sample.  The black points show direct
    constraints ($3\sigma$ limits) on the metallicity of a few galaxies via
    \Oiiib/\Hbeta, with 0.3\,dex uncertainty, using the direct method
    \citep{Curti2017}, conservatively assuming the upper branch solution for
    the upper limits.  The red star shows the average metallicity from the
    stack of the KMOS spectra (\autoref{fig:kmos_stack}) at the mean stellar
    mass (error bars correspond to the 16th and 84th percentile).  Overall, the
    measurements are consistent with the MZ relation at the typical redshift of
    the sample.\label{fig:MZalpha}}
\end{figure}
\begin{figure}[t]
  \includegraphics[width=\columnwidth]{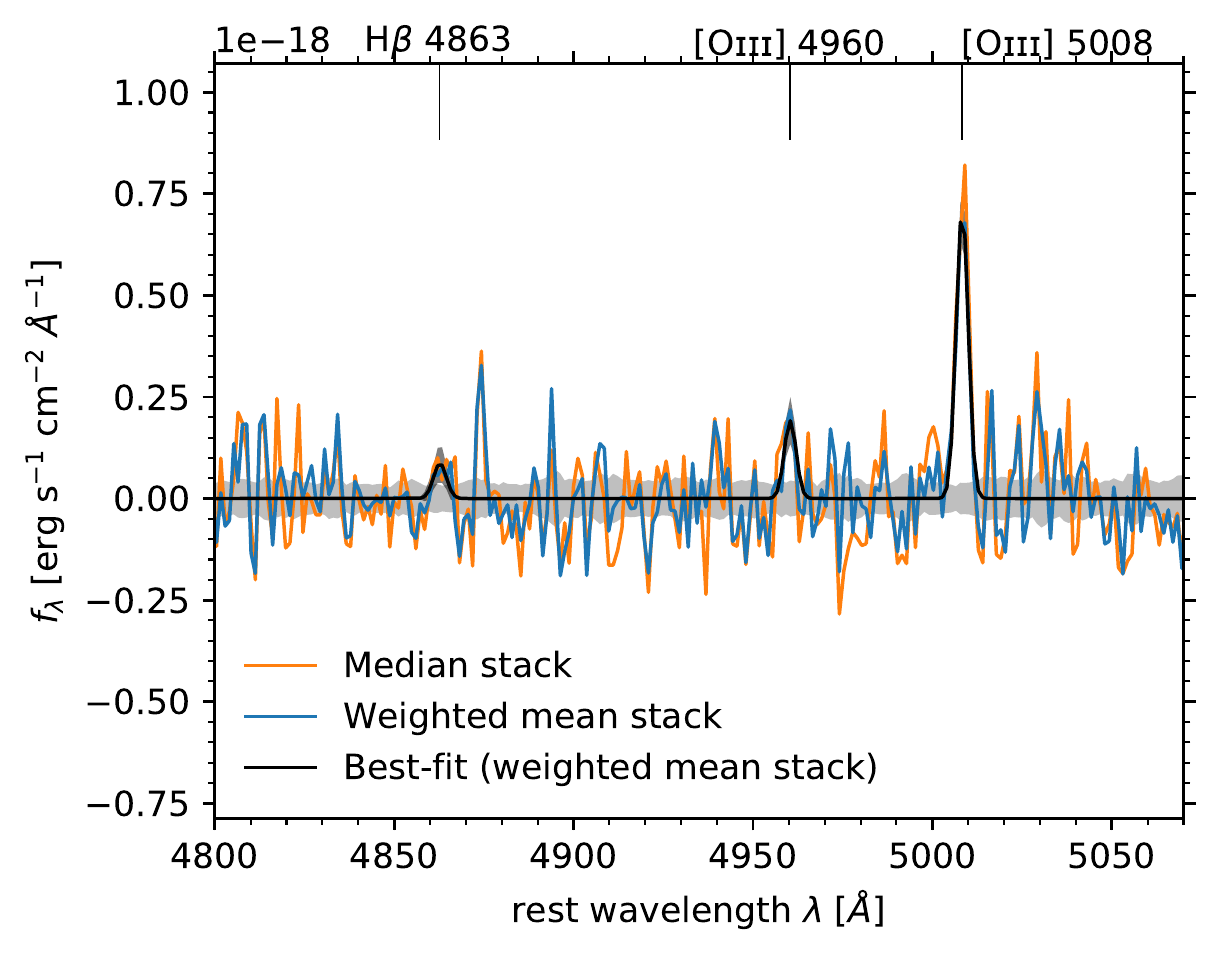}
  \caption{Stacked NIR spectra of the galaxies observed with KMOS (continuum
    subtracted).  We show the median stack (orange), the weighted mean stack
    (blue, with propagated errors in gray) and the best-fit to the weighted
    mean stack (in black, with uncertainties).  We tentatively detect \Hbeta,
    measuring $\log(\OIII/\Hb) \approx \logOIIIHbKMOSstack$, which broadly
    implies a metallicity of $\logOH \approx \logOHKMOSstack$
    \citep{Curti2017}.\label{fig:kmos_stack}}
\end{figure}
Stacking the $\avg{z} = \zavg$ star-forming galaxies in the HUDF by their
systemic redshifts, we find that
$L'_{\mathrm{CO}~J=4\rightarrow 3} \leq
\cofourstacklimitLp$\,K\,km\,s$^{-1}$\,pc$^{2}$.  This result puts an upper
limit on the molecular gas mass via
\begin{align}
  M_{\mathrm{mol}} = \aco r_{J1}^{-1} L'_{\mathrm{CO}~J\rightarrow J-1}, \label{eq:Mmol}
\end{align}
where $\aco$ is known as the CO-to-H$_{2}$ conversion factor (a light-to-mass
ratio) and $r_{J1}$ is the excitation correction, needed to convert the
observed CO luminosity to $L'_{\mathrm{CO}~J=1 \rightarrow 0 }$.  There is no
direct measurement of the CO excitation in the galaxies under consideration.
\cite{Valentino2020b} measured an average $r_{41} = 0.36 \pm 0.06$ in their
sample of star-forming galaxies at $z=1.25$, while \cite{Tacconi2018} assume an
average $r_{41} = 0.42$, constant with redshift.  \cite{Boogaard2020}, however,
have shown that there is significant evolution in the CO excitation of gas
mass-selected samples, with the average $r_{41}$ increasing from $0.3$ up to
$0.6$ between $z=1-2$ and $z=2-3$.  Indeed, \cite{Dessauges-Zavadsky2017} find
$r_{41} = 0.60 \pm 0.17$ in the strongly lensed MACSJ0032-arc at $z=3.6$, with
a similar $M_{*} \approx 10^{9.7}\,\Msun$ as our sample average (though
significantly higher $\mathrm{SFR} \approx 50$\,\Msun\,yr$^{-1}$).  If the
excitation scales with SFR surface density \citep{Daddi2015, Valentino2020b,
  Boogaard2020}, this may suggest the excitation to be lower in our galaxies on
average.  We therefore loosely assume that $r_{41} = 0.5 \pm 0.2$, broadly
encompassing the literature values.  Note that higher excitation implies a
smaller gas mass (\autoref{eq:Mmol}).  As we are dealing with an upper limit on
the gas mass in the first place, we effectively assume a lower limit on the
excitation of $r_{41} \ge 0.3$, in good agreement with observations.

A major uncertainty in the molecular gas estimate comes from \aco.  For star
forming galaxies at high redshift `Galactic' conversion factors are typically
assumed, consistent with observations in massive star forming galaxies
\citep{Daddi2010}.  However, the value of \aco\ has been observed to increase
strongly at low metallicity ($Z$; \citealt{Maloney1988, Israel1997}), where the
decreased shielding leads to dissociation of CO deeper into the clouds (e.g.,
\citealt{Wolfire2010}, see \citealt{Bolatto2013} for a review).  Several
calibrations for the metallicity dependence of the CO-to-H$_{2}$ conversion
factor exist in the literature, both determined empirically
\citep[e.g.,][]{Leroy2011, Magdis2012, Schruba2012, Genzel2012, Sandstrom2013}
as well as theoretically (e.g., \citealt{Wolfire2010}, see also
\citealt{Bolatto2013}).

We adopt the relation from \citet[][also adopted by \citealt{Tacconi2018,
  Dessauges-Zavadsky2020}]{Genzel2015},
\begin{align}
  \aco(Z)&=\aco^{\mathrm{MW}} \times \sqrt{10^{-1.27(\logOH -8.67)}} \\
         &\times \sqrt{0.67 \exp\left(0.36 \times 10^{-(\logOH-8.67)}\right)}, \label{eq:aco}
\end{align}
which is the geometrical mean of the curves from \citet[][which follows a
power-law, as the other literature relations do]{Genzel2012} and \citet[][which
has a steeper, exponential increase towards very low
metallicities]{Bolatto2013}.  Here, \logOH\ is the gas-phase oxygen abundance,
measured on the \cite{Pettini2004} scale (for conversion between metallicity
scales see \citealt{Kewley2008}), calibrated to a solar abundance of
$\logOH_{\odot} = 8.67$ \citep{Asplund2009}, and
$\aco^{\mathrm{MW}}=4.36$\,\Msun(K\,km\,s$^{-1}$\,pc$^{2}$)$^{-1}$, which
includes a factor 1.36 for helium \citep{Strong1996}.

To obtain metallicities in the absence of a direct tracer, the mass metallicity
relation can be used (see \autoref{sec:low-metall-drive} for discussion).
\cite{Genzel2015} determined the following mass-metallicity relation.
\begin{align}
  \logOH_{\mathrm{PP04}} = a - 0.087\left(\log{M_{*}} - b(z)\right)^{2}, \label{eq:MZ}
\end{align}
where $a=8.74 (0.06)$ and
$b(z) = 10.4\,(0.05) + 4.46\,(0.3) \log(1+z) - 1.78\,(0.4) \log^{2}(1+z)$ (uncertainties
in brackets), which is determined by combining several relations at different
redshifts \citep{Erb2006a, Maiolino2008, Zahid2014, Wuyts2014}.  Notably, this
relation approaches that of \cite{Maiolino2008} determined at $z=3.5$.

Alternatively, we can determine \Mmol\ from \CI, under the assumption of an
excitation temperature $\Tex$ and a neutral atomic carbon abundance
(\citealt{Weiss2005a}; see \citealt{Boogaard2020} for a detailed description).
We adopt $\Tex = 30$\,K \citep[][note that the atomic carbon mass is not a
strong function of excitation temperature above $\Tex=20$\,K]{Walter2011} and
an abundance of $\CI/[\mathrm{H}_{2}] = 2 \times 10^{-5}$ \citep{Valentino2018,
  Boogaard2020}.  We will revisit the latter assumption in
\autoref{sec:low-metall-drive}.

The dust can be used as a third tracer of the molecular gas mass.  We compute
the dust mass by relying on assumption that the Rayleigh-Jeans (RJ) tail of the
dust blackbody at long wavelengths is nearly always optically thin
\citep{Scoville2016}.  Specifically, we follow \cite{Magnelli2020} and assume a
dust opacity of $\kappa_{\nu_{0}} = 0.0431$\,m$^{2}$\,kg$^{-1}$ at
$\nu_{0} = 352.6$\,GHz (i.e., 850\,\micron; \citealt{Li2001}),\footnote{As
  pointed out by \cite{Magnelli2020}, assuming a typical gas-to-dust mass ratio
  of 100 (at solar metallicity), this dust mass absorption cross section is
  within a few percent of the ``ISM'' mass absorption cross section calibrated
  by \citet[][i.e., their $\alpha_{850\,\micron}$]{Scoville2016}.} a
mass-weighted mean dust temperature of $\avg{\Tdust}_{\mathrm{M}} = 25$\,K, and
a dust emissivity spectral index of $\beta=1.8$.  As argued by
\cite{Scoville2016}, the cold dust is the dominant contributor to the dust mass
and the RJ-tail of the dust emission, and recent studies by \emph{Planck} and
\emph{Herschel} have found the temperature to be in the range of 15--35\,K
\cite[e.g.,][]{PlanckCollaboration2011, Magnelli2014}.  Varying beta between
$1.5-2.0$ (the range typically assumed for the larger grains that dominate the
far-infrared emission, e.g., \citealt{DaCunha2008}) impacts the dust masses by
20--40\%.  Varying the dust temperature between 15--35\,K has a more
significant impact on the inferred gas masses, ranging from a factor 5.0 to
0.5, because the observations at rest-frame 275\,\micron\, start probing the
emission away from the RJ-tail and closer to the peak (this is further
discussed in \autoref{sec:low-metall-drive}).  We correct for the impact of the
Cosmic Microwave Background on the equilibrium dust temperature and observed
flux density \citep{DaCunha2013}, which increases the inferred mass by 10\%.

To convert the dust masses to gas masses, we assume a metallicity dependent
gas-to-dust ratio ($\gdr \simeq \Mmol/\Mdust$), with $\gdr(\Zsun) = 100$
\citep{Draine2007b}, making the common assumption that the gas in our galaxies
at $z=\zavg$ is predominately molecular \citep[e.g.,][]{Daddi2010a, Genzel2015,
  Tacconi2018}.  The $\gdr \propto Z^{\gamma}$ has been observed to decrease
close to linearly towards sub-solar metallicities \citep[with
$\gamma \approx -1$, e.g.,][]{Leroy2011, Magdis2012, Sandstrom2013,
  Saintonge2013}.  However, there is increasing evidence of a steeper relation
for metallicities below $\logOH \approx 8.0 - 8.1$; \cite{Remy-Ruyer2014} find
$\gamma \approx -3.1$ in local galaxies, while observations at $z\sim2$ suggest
that $\gamma < -2.2$ \citep{Coogan2019}, in agreement with the fiducial model
from \cite{Popping2017a}.  We explore both regimes, assuming the power-law
$\gdr(Z)$ relation from \cite{Tacconi2018} for a shallower increase with
metallicity.  For a steeper $\gdr(Z)$ at low metallicity, we adopt the broken
power law relation from \citet[][the $X_{\mathrm{CO},Z}$-case]{Remy-Ruyer2014},
which we scale to the same assumptions ($\logOH_{\odot} = 8.67$;
$\gdr(\Zsun)=100$):
\begin{align}
\gdr(Z) =
\begin{cases}
10^{2 + \gamma_{\mathrm{H}}(x - 8.67)} & \text{for } x > 8.1\\ \label{eq:gdr}
10^{0.8 + \gamma_{\mathrm{L}}(x - 8.67)} & \text{for } x \le 8.1,
\end{cases}
\end{align}
with $x = \logOH$.  Here $\gamma_{\mathrm{H}} = -1$ and
$\gamma_{\mathrm{L}} = -3.1$ are the power law slopes at high and low
metallicity, respectively.  The relation from \cite{Tacconi2018} is obtained
from \autoref{eq:gdr} by taking the $x > 8.1$ solution at all metallicities,
with $\gamma_{\mathrm{H}} = -0.85$.

\subsection{Low metallicity driving a high molecular gas mass-to-light ratio}
\label{sec:low-metall-drive}
We show the constraints on the molecular gas mass in the context of the
molecular gas-to-stellar mass-ratio ($\mugas = \Mmol/M_{*}$) and the gas
fraction ($\fgas = \Mmol/(\Mmol + M_{*})$), including a literature sample of CO
observations at low and high redshift, in \autoref{fig:fgas}.  At the basis of
the literature sample, we take the mass-selected sample of Sloan Digital Sky
Survey galaxies at $z=0$ from xCOLDGASS \citep{Saintonge2017}, together with
the massive, main-sequence selected galaxies at $z=0.5 - 2.5$ from the Plateau
de Bure HIgh-z Blue Sequence Survey (PHIBSS1+2) from the \cite{Tacconi2018}
compilation, and the galaxies from ASPECS at $z=1.0 - 3.6$ \citep{Aravena2019,
  Boogaard2019}.  We supplement these with studies that contain observations of
CO in (strongly lensed) star-forming galaxies at $z\ge3$ from \citet[using the
updated values from \citealt{Tan2013}]{Magdis2012, Magdis2017}, \citet[which
include two sources from
\citealt{Riechers2010}]{Dessauges-Zavadsky2015,Dessauges-Zavadsky2017},
\cite{Gowardhan2019} and \citet[based on the sources from
\citealt{Schinnerer2016}]{Cassata2020a}.  We convert the literature
observations to the metallicity dependent \aco\ (\autoref{eq:aco}; using the
mass metallicity relation when needed, \autoref{eq:MZ}) and adopt
$r_{21} = 0.77$ and $r_{31} = 0.55$ ($\pm 0.1$; to remain consistent with
\citealt{Tacconi2018}), though we keep the excitation corrections as assumed by
the authors in case these are better constrained through additional line
measurements \citep{Boogaard2020, Cassata2020a}.\footnote{For example, in the
  case of ASPECS, the measured $\avg{r_{31}} = 0.8$ \citep{Riechers2020b,
    Boogaard2020}, implies a factor 1.5$\times$ higher gas masses than the
  average value from \cite{Tacconi2018}.  Note however that, as argued in
  \autoref{sec:molecular-gas-masses}, differences in the excitation do not
  affect the upper limit on the gas mass of our star-forming galaxies at
  $\avg{z}=\zavg$, unless the excitation is significantly lower than our
  (conservative) lower limit.}

Assuming conversion factors that apply at solar metallicity
($\aco^{\mathrm{MW}} = 4.36$, $\gdr=100$), the stacking results imply gas
fractions that appear to be in tension with the observed gas fractions in
galaxies at $z\ge3$ at a similar stellar mass (see \autoref{fig:fgas}).  This
is in particular true for stringent limit based the dust, which places our
low-mass galaxies among the lowest gas fractions observed at $z=3-4$, with
$\fgas \leq 0.5$.  For CO, the tension becomes more clear once we take into
account that our galaxies are over an order of magnitude lower in stellar mass
than the typical galaxy studied in molecular gas at high redshift.  The gas
fraction in star-forming galaxies is observed to increase towards lower masses
and expected to be substantial for low-mass galaxies at these epochs
\citep{Scoville2017, Tacconi2018, Liu2019b}.  For reference, we show the
predicted gas fraction for a main sequence galaxy with
$M_{*} = 10^{9.1 \pm 0.2}$\,\Msun\ from \citet{Tacconi2018},\footnote{We adopt
  the ``$\beta = 0$'' scaling relation from \cite{Tacconi2018}, assuming a main
  sequence as observed by \citealt{Whitaker2014}.  Using their alternative
  ``$\beta=2$'' relation instead (which predicts a stronger increase in the gas
  fractions at lower redshift, with a turnover towards decreasing gas fractions
  above $z \sim 3$), the upper limits are still below the nominal value, but
  the limit based on CO falls within the scatter.}  taking into account an
extra 0.2 dex uncertainty in the average stellar mass, as well as for
$M_{*} = 10^{10}$\,\Msun\ (that is, more massive than the most massive galaxy
in our sample).  We also show the predicted gas fractions from \cite{Liu2019b},
based on dust continuum measurements only, which are higher than those from
\citet[][the relations from \citealt{Scoville2017} predict even higher gas
fractions]{Tacconi2018}.  Taking into account the evolution of the gas fraction
in low-mass galaxies, the upper limit based on the CO is also in tension with
the expected gas fraction.

At face value, this result suggests that the galaxies in our sample have
unexpectedly low molecular gas fractions.  However, a more likely explanation
is that the assumption of a Galactic \aco\ and gas-to-dust ratio does not hold
for these systems.  Indeed, significantly higher conversion factors would be
naturally explained by sub-solar metallicities for these systems.

In \autoref{fig:MZalpha}, we show the MZ relation from \autoref{eq:MZ} at the
average redshift of our sample.  We find that the metallicity at the median
mass (16th, 84th percentile) of the sample is $\logOH = \logOHmed$.  However,
the MZ relation is only an approximate tracer of the metallicity.  More
directly, the $\Oiiib/\Hbeta$-ratio can be used to trace the metallicity,
albeit with significant scatter, as the ratio monotonically increases with
decreasing metallicity, up to a turnover at $\logOH \sim 8.0$
\citep[e.g.,][]{Curti2017, Sanders2020}.  Because of this turnover, there are
two metallicities solutions at a fixed ratio; one on the upper branch (high
metallicity) and one on the lower branch (low metallicity).  We robustly detect
\Hb\ in two objects, finding high ratios of $\log(\OIII/\Hb) \approx 0.8$ for
two (MUSE-1019 and MUSE-6878).  This roughly implies a metallicity at the
turnover, $\logOH \sim 8.0$, via the direct method
\citep{Curti2017}.\footnote{http://www.arcetri.astro.it/metallicity/calibrazioni.pl}
We also tentatively detect $\Hbeta$ in a third object (MUSE-6895), yielding a
lower ratio ($0.35$), which implies a high metallicity if it is on the upper
branch of the metallicity calibration, in tension with its stellar mass.
However, assuming that it follows the (extrapolated) lower branch, this would
imply a much lower metallicity of $\logOH \sim 7.35$ (and an extremely high
$\aco \gg 100$), in better agreement with the stellar mass.  For the remaining
galaxies, we only find (weak) upper limits on the metallicity (conservatively
assuming all are on the higher branch).

To obtain an estimate of the average metallicity in the sample, we stack the
KMOS spectra using weighted mean and median stacking.  Because of the
uncertainties in the background level (see \autoref{sec:kmos}), we do not
normalize the spectra but stack the continuum subtracted spectra instead, which
may introduce a bias towards the brighter objects that go into the stack.  Note
that, due to the shifting to a common redshift, the skyline residuals are
spread throughout the stack, though this problem should be mitigated in the
median stack.  The stacked spectra are shown in \autoref{fig:kmos_stack}.  We
tentatively detect \Hbeta\ at 2--3$\sigma$, measuring
$\log(\OIII/\Hb) \approx \logOIIIHbKMOSstack$, which broadly implies a
metallicity of $\logOH \approx \logOHKMOSstack$.  We do caution against
over-interpreting the stack, given the uncertainties mentioned above.  It
should also be stressed that the mass-metallicity relation only holds on
average.  For example, \cite{Dessauges-Zavadsky2017}, found that the lensed arc
at $z=3.6$ has a higher inferred metallicity from its measured \OIII/\Hb-ratio
than predicted from the MZ.  While the sample selection could in principle bias
the average metallicity, this is not immediately obvious, as the selection
towards galaxies with a low EW(\Lya) but a high EW(\OIII) bias the metallicity
in opposite directions and may to some extent cancel out.  Overall, the
metallicities of the individual galaxies and the average metallicity from the
stack are in reasonable agreement with the predictions from the MZ relation
(\autoref{fig:MZalpha}), pointing to an average metallicity of
$\logOH \approx \logOHavg$ for the galaxies in our sample.

The average metallicity of our sample implies a significantly higher value of
$\aco \approx \acoavg$, which places our upper limit in comfortable agreement
with the predicted gas fractions (see \autoref{fig:fgas}).  Notice that the
strong, non-linear increase in the conversion factor with metallicity
(\autoref{eq:aco}) makes the exact value uncertain, particularly in the
low-mass range.  Furthermore, we caution that there is still debate about the
exact relation between \aco\ and metallicity at low metallicity, mostly due to
the difficulty of constraining \aco\ at low metallicity.  In any case, a
minimal value of $\aco \geq \acomin$ is required to place the $3\sigma$ upper
limit on the \cite{Tacconi2018} relation, more than 2 times the Galactic value.

The shallower relations between gas-to-dust ratio and metallicity yield
gas-to-dust ratios that are insufficient to reconcile the observed limit with
the scaling relations, which requires a $\gdr \geq \gdrmin$.  For example, we
find $\gdr \approx \gdravglin$ based on \cite[][see
\autoref{eq:gdr}]{Tacconi2018}.  This points towards a steeper relation between
the \gdr\ and metallicity in the low-metallicity regime, as suggested by, for
example, \cite{Remy-Ruyer2014} and \cite{Coogan2019} (see
\autoref{sec:molecular-gas-masses}).  Adopting the relation from
\citet[][\autoref{eq:gdr}]{Remy-Ruyer2014} yields a significantly higher
$\gdr \approx \gdravgbpl$, again placing our upper limit in comfortable
agreement with the expected gas fraction.  Alternatively, a dust temperature of
$\Tdust < 20$\,K (as opposed to 25\,K), can also reconcile the difference with
a lower $\gdr$.  However, given the stronger radiation field expected at low
metallicity, it is not clear that the dust temperatures would be much lower
than assumed.

Independent of CO and the dust, we reach similar conclusions using \CI\
instead.  While a typical abundance ratio of $2\times10^{-5}$ predicts a gas
mass that is marginally consistent with the scaling relations, assuming an
lower abundance (e.g., $ 8\times$; loosely based on the metallicity) easily
yields a limit that is fully consistent with the scaling relations, with a
$\approx 2 \times$ lower abundance being minimally required (see
\autoref{fig:fgas}).  Note that the metallicity impacts CO and \CI\ in
different ways, however, such that \CI\ may be a preferred over CO as tracer of
cold gas in low metallicity environments (we will come back to this point in
\autoref{sec:impl-obta-gas}).

For reference, we collect the different molecular gas masses and gas-to-stellar
mass ratios that are discussed in this section, and shown in
\autoref{fig:fgas}, in \autoref{tab:mgas-results}.

\begin{deluxetable}{lccc}
  \tablecaption{Upper limits ($3\sigma$) on the molecular gas mass and
    molecular gas-to-stellar mass ratio under different
    assumptions.\label{tab:mgas-results}}

    \tablehead{
    \colhead{Tracer} &
    \colhead{Conversion factor} &
    \colhead{$\avg{\Mmol}$} &
    \colhead{$\avg{\mugas}$}\\
    \colhead{} &
    \colhead{} &
    \colhead{($\times 10^{9}$\,\Msun)} &
    \colhead{}
  }
  \colnumbers
  \startdata
 \multicolumn{4}{c}{Solar metallicity}\\
 \hline
 CO $J=4\rightarrow3$     & $\aco = 4.36 $ & $< \colimitmmolsolar$ & $< \colimitmusolar$ \\
 \CIthreePone             & $X_{\CI} = 2\times 10^{-5} $ & $< \cilimitmmolsolar$ & $< \cilimitmusolar$ \\
 $S_{\nu}$(1.2\,mm) & $\gdr = 100 $ & $< \dustlimitmmolsolar$ & $< \dustlimitmusolar$ \\
\hline
 \multicolumn{4}{c}{Sub-solar metallicity}\\
 \hline
 CO $J=4\rightarrow3$     & $\aco = \acoavg $ & $< \colimitmmolsubsolar$ & $< \colimitmusubsolar$ \\
 \CIthreePone             & $X_{\CI} = 2.5\times 10^{-6}$ & $< \cilimitmmolsubsolar$ & $< \cilimitmusubsolar$ \\
 $S_{\nu}$(1.2\,mm) & $\gdr = \gdravglin $ & $< \dustlimitmmolsubsolarlin$ & $< \dustlimitmusubsolarlin$ \\
 $S_{\nu}$(1.2\,mm) & $\gdr = \gdravgbpl $ & $< \dustlimitmmolsubsolarbpl$ & $< \dustlimitmusubsolarbpl$
 \enddata

 \tablecomments{Derived from the upper limits (see \autoref{sec:stacking} and
   \autoref{tab:lines}) for the full systemic redshift sample
   ($\avg{M_{*}} = 10^{9.1}$\,\Msun) as explained in
   \autoref{sec:molecular-gas-masses}, with metallicity dependent conversion
   factors as listed.  (1) Molecular gas tracer.  (2) Adopted (metallicity
   dependent) conversion factor ($X_{\CI} \equiv \CI / [\mathrm{\HH}]$). (3)
   Upper limit on molecular gas mass. (4) Upper limit on
   $\avg{\mugas} = \avg{\Mmol} / \avg{M_{*}}$.}
\end{deluxetable}
\subsection{Contribution to the cosmic molecular gas density}
\label{sec:maxim-contr-cosm}
\begin{figure}[t]
  \centering
  \includegraphics[width=\columnwidth]{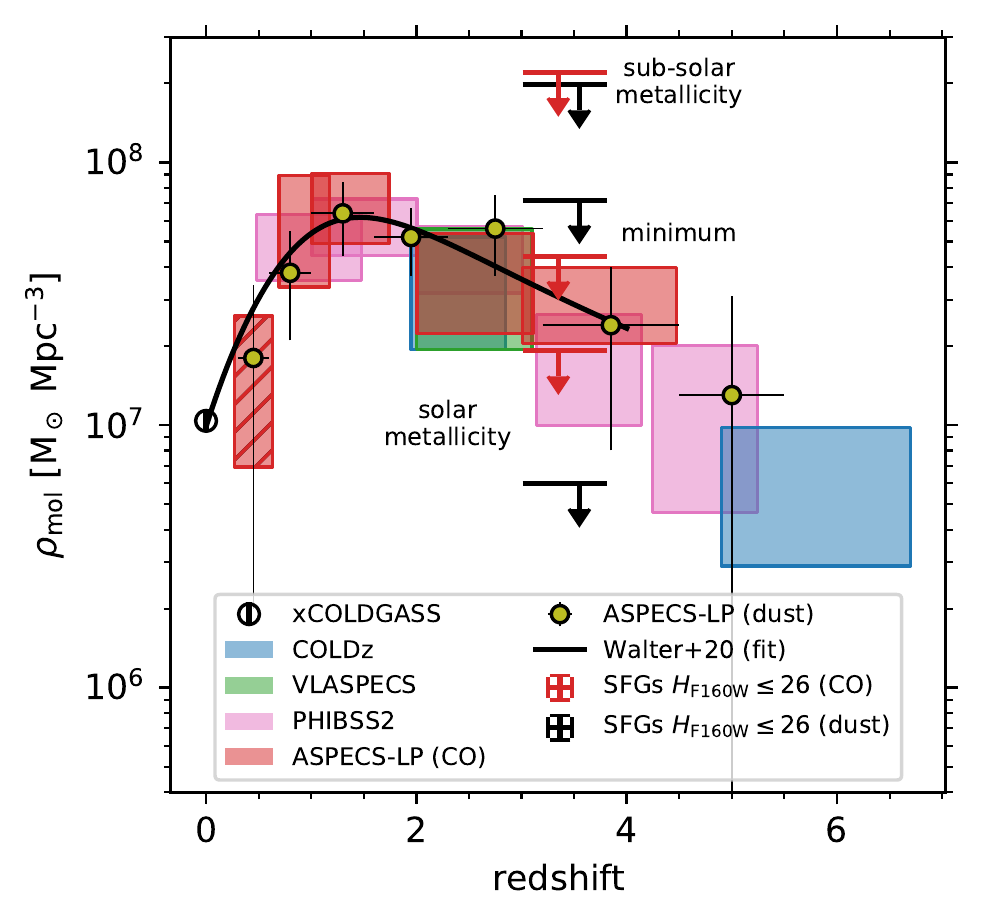}
  \caption{Cosmic molecular gas density as a function of redshift.  The
    literature data is from \cite{Fletcher2020} (xCOLDGASS),
    \cite{Riechers2019} (COLDz), \cite{Riechers2020b} (VLASPECS),
    \cite{Lenkic2020} (PHIBSS2), \cite{Decarli2020} (ASPECS-LP CO),
    \cite{Magnelli2020} (ASPECS-LP dust) and we also show the best-fit from
    \cite{Walter2020}.  We show the estimates on the upper limit on the cosmic
    molecular density for all galaxies with $\Hmag \leq 26$ at
    $3.0115 < z < 3.812$, as derived from CO (red; assuming $r_{41} = 0.5$) and
    the dust-continuum (black), under the assumption our stacked averages are
    representative for the larger population.  The different limits are for
    conversion factors $(\aco, \gdr)$ = (4.36, 100), (\acomin, \gdrmin),
    (\acoavg, \gdravgbpl), corresponding to solar metallicity, the minimum
    value based on \autoref{sec:low-metall-drive}, and the best-estimate
    sub-solar metallicity (the CO-based literature data is scaled to match our
    assumption on $\aco^{\rm MW}$).  The upper limits do not rule out a large
    amount of molecular gas in lower mass galaxies (that would have been missed
    in previous surveys, due to their high gas mass-to-light ratio), but are
    equally consistent with a smaller contribution to the total molecular gas
    budget.  \label{fig:rho-mol}}
\end{figure}
The galaxies under study are below the detection threshold of current
$\rhomol(z)$ surveys \citep[e.g.,][]{Riechers2019, Decarli2019, Decarli2020}.
Still, their potentially high gas mass-to-light ratios imply that they could
have a significant contribution to the total cosmic molecular gas density.
Assuming the average gas masses derived from the stacks are representative of
all 67 galaxies in the photometric catalog down to $\Hmag = 26$ (cf.
\autoref{fig:z-f160}), we compute the total contribution of these galaxies to
the cosmic molecular gas density, $\rho(3.0115 < z < 3.812)$.  We adopt the
solar, the minimum, and the sub-solar conversion factors from
\autoref{sec:low-metall-drive}, and $r_{41}=0.5$, and shift the CO-based
determinations of $\rhomol$ at $z>0$ to match our assumption on
$\aco^{\rm MW}$.  The result can be seen in \autoref{fig:rho-mol}.  Because the
upper limits are not stringent enough, the results are inconclusive.  On one
hand, they do not exclude the possibility that a significant amount of
molecular gas is missed due to the high gas mass-to-light ratio of star forming
galaxies at these redshifts.  On the other hand, it is equally possible that
their contribution is significantly smaller, implying that their molecular gas
signal lies well below the detection threshold, even in stacks.  We caution
that the strong increase in the conversion factor with decreasing metallicity
(especially for the lowest metallicity sources) is a significant source of
uncertainty when extrapolating the averages to sources over a larger range in
mass and metallicity.  We therefore also determine upper limits for the more
massive galaxies in the sample only (with $M_{*} \ge 10^{9}$\,\Msun\ and
$\ge 10^{9.5}$\,\Msun, computing their average conversion factors from the
mass-metallicity relation).  The limits on \rhomol\ are slightly stronger for
these sub-samples because, while the stacks are slightly less constraining, the
estimated \aco\ and \gdr\ are lower, as well as the number of sources in the
volume.  The results fall in between the minimum and sub-solar values of
magnitude limited sample, but do not alter the conclusions overall

\subsection{Implications for observing molecular gas in low metallicity galaxies at high redshift}
\label{sec:impl-obta-gas}
The evolution of the metallicity of star-forming galaxies with redshift has
significant implications for the detectability of molecular gas at $z\ge3$.
Even in the local universe, detecting CO in low metallicity dwarf galaxies has
been challenging \citep[e.g.,][]{Schruba2012, Hunt2015}.  The substantial
CO-to-H$_{2}$ conversion factor and gas-to-dust ratios inferred for our
low-mass, low metallicity galaxies imply that detecting the molecular gas
reservoir in these galaxies will be very challenging on an individual basis,
even with modern instruments.  Similar conclusions are also reached for more
massive galaxies at sub-solar metallicities \citep[e.g.][]{Tan2013,
  Coogan2019}.  \cite{Tan2013} have shown that under the assumption of a
MZ-\aco\ relation similar to the one adopted here, the expected CO luminosity
for a star-forming galaxy on the main sequence rapidly declines, due to the
metallicity evolution.  This raises the interesting question of how the
molecular gas content can be best constrained in sub-solar metallicity
star-forming galaxies at high redshift.

There are significant uncertainties in deriving a total dust and gas mass from
the dust continuum in the low metallicity regime.  Variations in the process
and balance of dust formation and destruction at low metallicity, as well as
differences in grain composition and size distribution can have a major impact
on the gas-to-dust ratio, the dust emissivity and emerging dust spectrum (e.g.,
\citealt{Remy-Ruyer2014}, see also \citealt{Draine2007b}).  In addition to
these complications, the fainter part Rayleigh-Jeans tail at long wavelengths
has to be probed, such that the blackbody is dominated by cold dust which
dominates the mass and the uncertainty in the (unknown) mass-weighted
dust-temperature is minimized \citep[e.g.][]{Scoville2016}.

At low metallicity, CO also becomes an increasingly poor tracer of the total
molecular gas reservoir.  Because of the lower dust abundance at low
metallicity, CO is dissociated and ionized into C and C$^{+}$ deeper into the
clouds, while the H$_{2}$ self-shields against photodissociation
\citep{Gnedin2014}, resulting in an increasing volume of H$_{2}$ gas that is
not traced by CO at low metallicity \citep[e.g.,][]{Wolfire2010}.  This
provides motivation to investigate and develop the theoretical underpinning for
other species as tracers of the molecular gas, such as the fainter \CI\ lines
(e.g., \citealt{Weiss2003, Weiss2005a, Papadopoulos2004}; see also
\citealt{Valentino2018, Boogaard2020}), but in particular also \CIIfsl.

The bright \CIIonehalfP\ line at 158\,\micron\ is one of the foremost cooling
lines of the ISM, also at low metallicity \citep[where it is outranked only by
the high ionization line \OIIIfslb\ at 35\,eV; e.g.,][]{Cormier2015,
  Cormier2019} and its high luminosity allows it to be observed in star-forming
galaxies out to the highest redshifts \citep{Ouchi2013, Ota2014, Maiolino2015,
  Capak2015, Knudsen2016, Pentericci2016, Bradac2017, Matthee2017, Matthee2019,
  Carniani2018a, Carniani2018b, Carniani2020, Smit2018, Hashimoto2018,
  Laporte2019, LeFevre2019, Bethermin2020, Harikane2020a, Bakx2020}.  With an
ionization potential of 11.3\,eV (that is, lower than \HI\ at 13.6\,eV), \CII\
can arise in both the neutral and ionized medium, though it becomes an
increasingly better tracer of the neutral ISM towards lower metallicities
(\citealt{Croxall2017, Cormier2019}, see also \citealt{Diaz-Santos2017})
potentially due to the carbon in the \HII-regions being further ionized into
C$^{++}$ (as witnessed by the shift in the ionization balance to high
ionization lines, also see in some high-$z$ sources, e.g., \citealt{Pavesi2016,
  Harikane2020a}, see also \citealt{Carniani2020}).  While, as a cooling line,
\CII\ is in principal sensitive to the heating rate (and not the molecular gas
mass), it may be calibrated as a molecular gas tracer \citep[][]{Zanella2018}.
As such, \CII\ can potentially outperform other tracers of the molecular gas
mass, in particular in low metallicity environments \cite[e.g.][]{Madden2020}.
As \CII\ is already seeing use as tracer of the molecular gas at high redshift
\citep[e.g.,][]{Dessauges-Zavadsky2020}, its use to this end should be further
investigated both observationally and theoretically.

\section{Summary and conclusions}
\label{sec:summary-conclusions}
We present constraints on the molecular gas signal for a sample of \Nobj\
star-forming galaxies at $\avg{z}=\zavg$, with a median $M_{*} = \Mmed$, in the
\emph{Hubble} Ultra Deep Field (HUDF).  Based on their \Lya- and
\Hmag-selection (\autoref{fig:z-f160}), they show relatively low
$\EWLya \approx \EWavg$\,\AA\ ($L_{\Lya} = 0.2\,L_{\Lya}^{*}$), and rest-frame UV
spectra similar to star-forming galaxies at the same epoch (see
\autoref{fig:muse_stack}).  We efficiently follow-up \Lya-selected galaxies
from the MUSE HUDF Survey, with near-infrared spectroscopy from KMOS and
MOSFIRE to determine their systemic redshifts (\autoref{fig:spectrum_example}
and \autoref{fig:spectra}) and stack the molecular line emission from the ALMA
Spectroscopic Survey in the HUDF (ASPECS).  Our main results are as follows:

\begin{itemize}
\item We determine systemic redshifts from the rest-frame UV and rest-frame
  optical features, finding an average velocity offset of
  $\avg{\Delta v(\Lya)} = \deltavlya$~km\,s$^{-1}$ (with a 100 to
  600\,km\,s$^{-1}$ range) consistent with the relatively low \EWLya\ of the
  sample (\autoref{fig:vel_offs}).
\item Stacking the signal from $^{12}\mathrm{CO}\ J=4\rightarrow3$ and
  \CIthreePone\ (as well as the $^{12}\mathrm{CO}\ J=9\rightarrow8$ and
  $J=10\rightarrow9$ lines), we do not find any detections and determine
  $3\sigma$ upper limits on the line luminosities of
  \cofourstacklimitLp~K\,km\,s$^{-1}$pc$^{2}$ and
  \cistacklimitLp~K\,km\,s$^{-1}$pc$^{2}$, respectively, for a linewidth of
  \deltavstack~km\,s$^{-1}$ (see \autoref{fig:stacks_all} and
  \autoref{tab:lines}; also for the limits on the higher-$J$ lines).  Stacking
  the dust continuum at 1.2\,mm and 3\,mm, we find $3\sigma$ upper limits on
  the flux densities of $S_{\nu} \leq \dustlimit$ and $\leq \dustlimitThreemm$,
  respectively (\autoref{fig:duststack}).
\item Comparing the inferred molecular fraction from CO and the dust continuum
  to scaling relations, we find that assuming a `Galactic`
  $\aco^{\mathrm{MW}} = 4.36$ and $\gdr = 100$ significantly underpredicts the
  expected molecular gas mass (\autoref{fig:fgas}).  In order to reconcile our
  measurements with the published scaling relations from \cite{Tacconi2018}
  would require an $\aco \geq \acomin$ and $\gdr \geq \gdrmin$.  This result
  either implies that our galaxies have unexpectedly low gas fractions or that
  the assumption of solar-metallicity conversion factors break down.
\item Using the mass-metallicity relation, as well as constraints from
  \Oiiib/\Hbeta\ (\autoref{fig:kmos_stack}), we predict an average metallicity
  of our sample of $\logOH = \logOHavg$, that is, significantly sub-solar.
  This implies a high $\aco \approx \acoavg$ making our result consistent with
  the expected (high) gas fractions at $z=3.5$ (\autoref{fig:MZalpha}).
\item An approximately linear scaling relation between the gas-to-dust ratio
  and metallicity \citep[$\gdr \propto \gamma^{-0.85}$;][]{Tacconi2018} yields
  $\gdr \approx \gdravglin$, which is insufficient to reconcile the limit based
  on the dust with the scaling relations.  Using a steeper relation at low
  metallicity \citep[$\gdr \propto \gamma^{-3.1}$ at
  $\logOH \leq 8.1$;][]{Remy-Ruyer2014} instead yields
  $\gdr \approx \gdravgbpl$, making our upper limit consistent again
  (\autoref{fig:fgas}).
\item Independent of the CO, we find a similar tension from the \CI\
  luminosity, which implies a $\CI/[\mathrm{H}_{2}]$ abundance lower than in
  massive star-forming galaxies such as the Milky Way (\autoref{fig:fgas}).
\item Based on the results, we compute the upper limit on the contribution of
  all galaxies with $\Hmag \leq 26$ to the cosmic molecular gas density
  $\rhomol(z=3.0115-3.812)$.  The upper limits are not constraining enough to
  exclude the possibility of a significant contribution from these galaxies,
  that lie below the detection threshold of current surveys, to the cosmic
  molecular gas density.
\end{itemize}

The results of this work exemplify the difficulty to obtain molecular gas mass
estimates in low metallicity environments, which are expected to be more
prevalent in typical star forming galaxies at $z\geq3$.  Given the
uncertainties associated with the dust and CO at low metallicity we argue for
the further observational and theoretical development of alternative tracers of
the molecular gas reservoir, such as the bright \CIIfsl\ line, that should be
more easily observable with ALMA.  Obtaining accurate constraints on the
gas-phase metallicity of high-redshift galaxies will key in this regard and one
of the key pieces of information that the \emph{James Webb Space Telescope}
will be able to provide.

\acknowledgments
We would like to thank the referee for a constructive and helpful
report. L.A.B. is grateful to Corentin Schreiber for assisting with the
near-infrared spectroscopy during the early stages of this work.
L.A.B. acknowledges support from the Leids Kerkhoven-Bosscha Fonds under
subsidy numbers 18.2.074 and 19.1.147.  D.R. acknowledges support from the
National Science Foundation under grant numbers AST-1614213 and
AST-1910107. D.R. also acknowledges support from the Alexander von Humboldt
Foundation through a Humboldt Research Fellowship for Experienced Researchers.
A.F. acknowledges the support from grant PRIN MIUR 2017-20173ML3WW\_001.
J.B. acknowledges support by Funda\c{c}{\~a}o para a Ci{\^e}cia e a Tecnologia
(FCT) through the research grants UID/FIS/04434/2019, UIDB/04434/2020,
UIDP/04434/2020.  H.I. acknowledges support from JSPS KAKENHI Grant Number
JP19K23462.

This work is based on observations collected at the European Southern
Observatory under ESO programs 094.A-2089(B), 095.A-0010(A), 096.A-0045(A),
096.A-0045(B), 099.A-0858(A), and 0101.A-0725(A).
This paper makes use of the following ALMA data: ADS/JAO.ALMA\#2016.1.00324.L.
ALMA is a partnership of ESO (representing its member states), NSF (USA) and
NINS (Japan), together with NRC (Canada), NSC and ASIAA (Taiwan), and KASI
(Republic of Korea), in cooperation with the Republic of Chile. The Joint ALMA
Observatory is operated by ESO, AUI/NRAO and NAOJ. The National Radio Astronomy
Observatory is a facility of the National Science Foundation operated under
cooperative agreement by Associated Universities, Inc.
This work was supported by a NASA Keck PI Data Award, administered by the NASA
Exoplanet Science Institute. Data presented herein were obtained at the
W. M. Keck Observatory from telescope time allocated to the National
Aeronautics and Space Administration through the agency's scientific
partnership with the California Institute of Technology and the University of
California. The Observatory was made possible by the generous financial support
of the W. M. Keck Foundation.  The authors wish to recognize and acknowledge
the very significant cultural role and reverence that the summit of Maunakea
has always had within the indigenous Hawaiian community.  We are most fortunate
to have the opportunity to conduct observations from this mountain.

\facilities{ALMA, VLT:Yepun (MUSE), VLT:Antu (KMOS), Keck I (MOSFIRE).}
\software{\textsc{Topcat} \citep{Taylor2005}, \textsc{Gnuastro}
  \citep{Akhlaghi2015}, \textsc{IPython} \citep{Perez2007}, \textsc{numpy}
  \citep{VanDerWalt2011}, \textsc{Matplotlib} \citep{Hunter2007},
  \textsc{Astropy} \citep{Robitaille2013, TheAstropyCollaboration2018},
  \textsc{Casa} \citep{McMullin2007}.}

\appendix
\section{Spectra}
\label{sec:spectra}
The rest-frame UV (MUSE) and near-IR (KMOS, MOSFIRE) spectra of all galaxies in
the sample are shown in \autoref{fig:spectra} (except for the galaxy
already shown in \autoref{fig:spectrum_example}).
\begin{figure*}[t]
  \centering
  \includegraphics[width=0.9\textwidth]{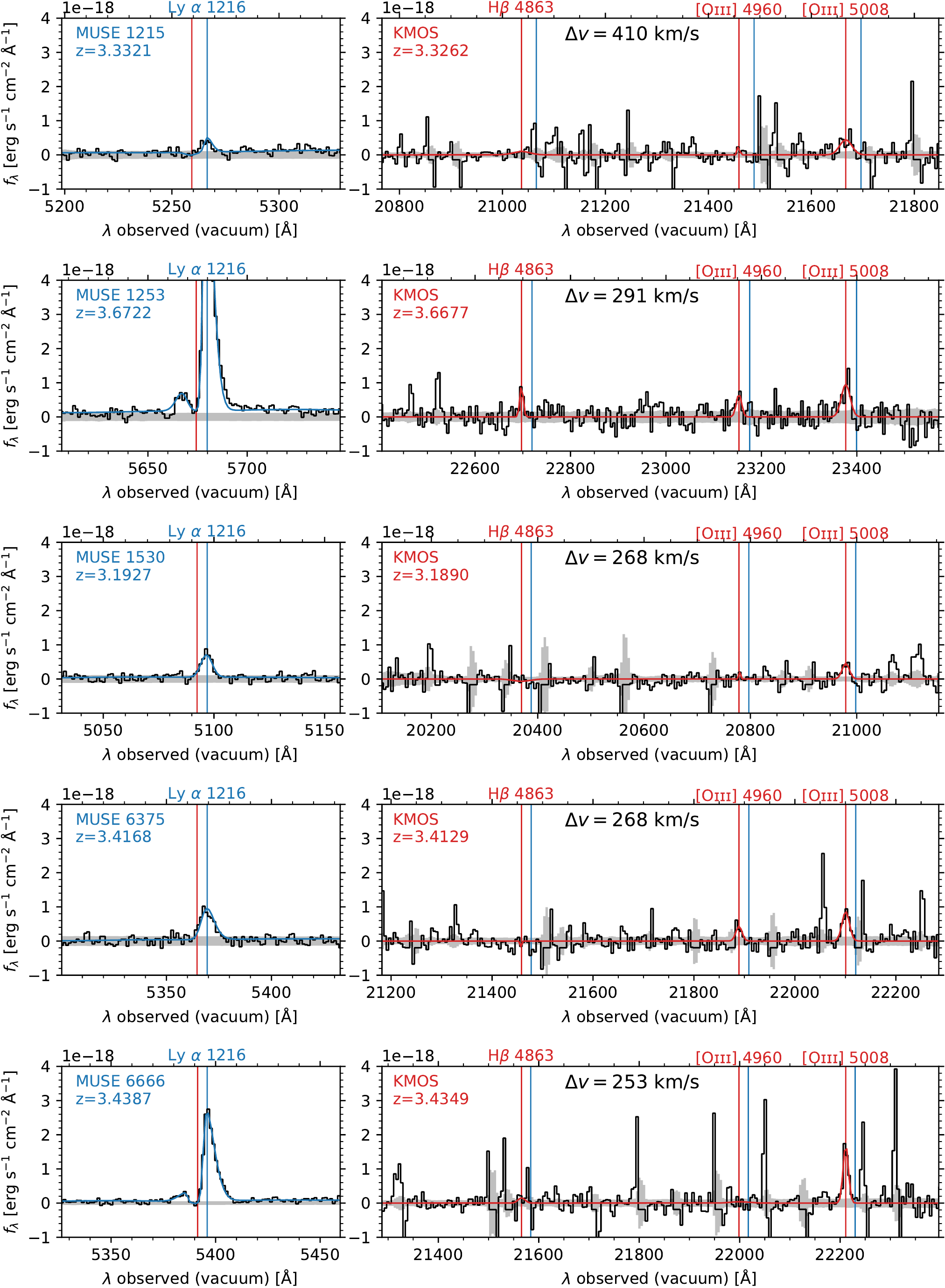}
  \caption{Rest-frame ultraviolet and optical spectra for the galaxies with
    near-infrared follow up.  The \textbf{left} panel shows the MUSE spectrum
    surrounding the \Lya\ line.  The \textbf{right} panel shows the continuum
    subtracted KMOS or MOSFIRE spectrum (indicated in the figure) around the
    \Hb\ and \OIII\ lines.  In both panels the vertical blue and red lines
    indicate the redshift of \Lya\ and the systemic redshift, respectively,
    determined from the fit to the spectrum (shown in the same color).  All
    spectra show a positive velocity offset between the red peak of \Lya\ and
    the systemic redshift. \label{fig:spectra}}
\end{figure*}
\begin{figure*}[t]
  \figurenum{\ref{fig:spectra}}
  \centering
  \includegraphics[width=0.9\textwidth]{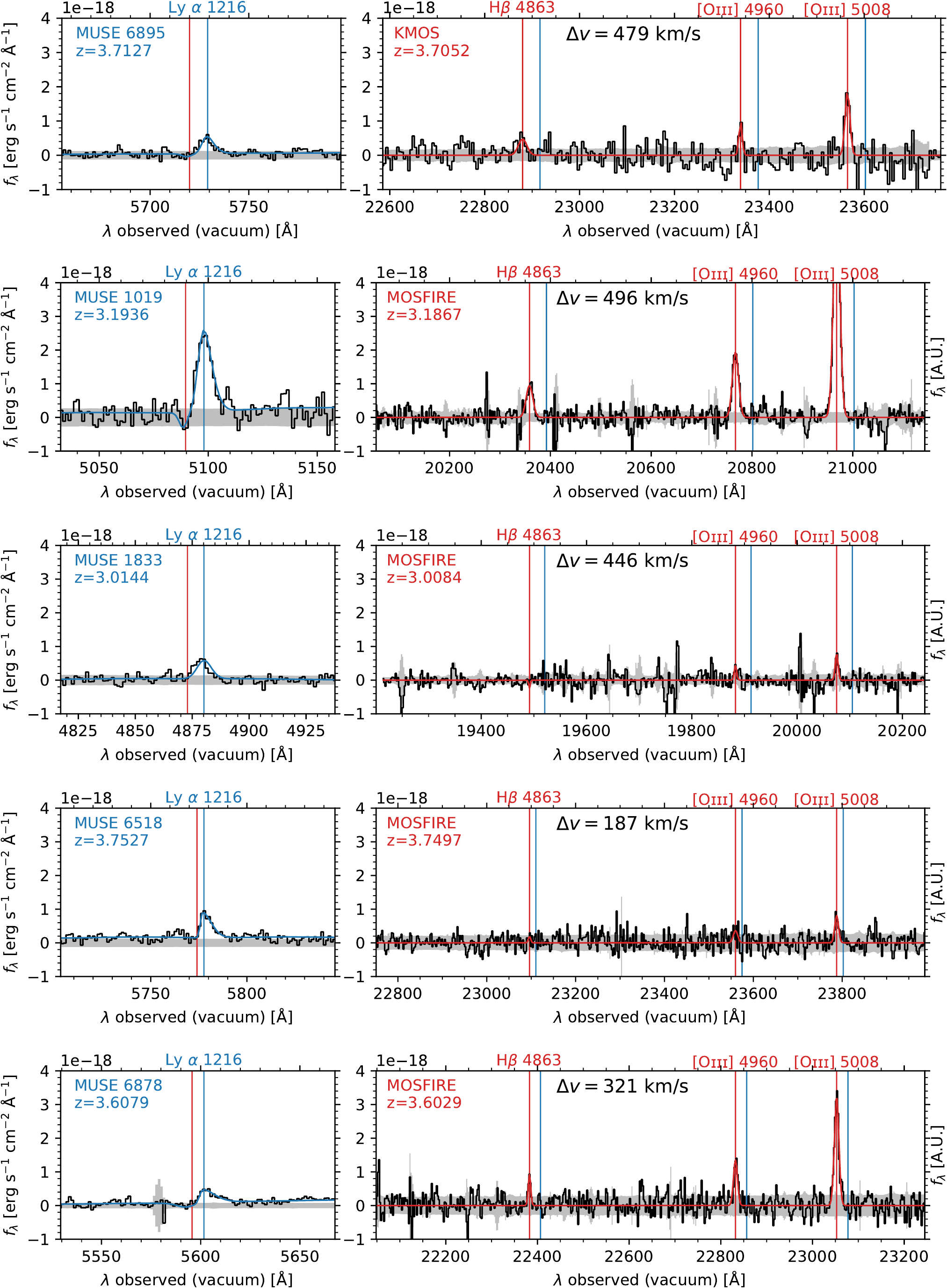}
  \caption{\emph{(continued)}}
\end{figure*}
\section{Table}
  The coordinates, (systemic) redshifts, and physical properties of the sample
  of $3.0115 < z < 3.812$ star-forming galaxies are listed in
  \autoref{tab:sources}.
\begin{longrotatetable}
    \movetabledown=5mm
  \begin{deluxetable*}{ccccCcCccCCC}
    \tablewidth{\textwidth}
  \tablecaption{Coordinates, (systemic) redshifts, and physical properties
      of the sample of $3.0115 < z < 3.812$ star-forming galaxies. \label{tab:sources}}
  \tablehead{
    \colhead{ID} & \colhead{RAF ID} & \colhead{$\alpha_{J2000}$} &
    \colhead{$\delta_{J2000}$} &
    \colhead{$\zLya$}&\colhead{$\delta(\zLya)$} &\colhead{\zsys} &
    \colhead{$\delta(\zsys)$} & \colhead{Source} & \colhead{$\Delta v
      (\Lya)$} & \colhead{$\log M_{*}$} & \colhead{log SFR}\\
    \colhead {}  & \colhead {}      & \colhead{}                & \colhead{}
    & \colhead{}  &   \colhead{(km\,s$^{-1}$)} & \colhead{} &
    \colhead{(km\,s$^{-1}$)} & & \colhead{(km\,s$^{-1}$)} & \colhead{(\Msun)} &
    \colhead{(\Msun\,yr$^{-1}$)}
  }
  \colnumbers
  \startdata
50               & 9110    & 53.16284897 & -27.77162645 & $3.33015 \pm 0.00086$ & 59.5    & $3.32349 \pm 0.00027$ & 18.7   & MUSE EM  & $461.8 \pm 62.5$   & $9.15_{-0.12}^{+0.10}$ & $1.122_{-0.10}^{+0.10}$ \\
82               & 6627    & 53.1515512  & -27.7853475  & $3.60777 \pm 0.00006$ & 3.9     & $3.60493 \pm 0.00080$ & 52.1   & MUSE EM  & $184.9 \pm 52.3$   & $9.00_{-0.10}^{+0.10}$ & $0.792_{-0.17}^{+0.10}$ \\
106              & 9863    & 53.16372638 & -27.77907551 & $3.28171 \pm 0.00009$ & 6.3     & $3.27648 \pm 0.00012$ & 8.4    & MUSE EM  & $366.6 \pm 10.5$   & $8.45_{-0.10}^{+0.10}$ & $0.547_{-0.10}^{+0.10}$ \\
118              & 23839   & 53.15708801 & -27.78026883 & $3.02127 \pm 0.00056$ & 41.7    & $3.01727 \pm 0.00017$ & 12.7   & MUSE EM  & $298.5 \pm 43.7$   & $8.78_{-0.20}^{+0.10}$ & $0.462_{-0.10}^{+0.10}$ \\
1019             & 8002    & 53.16492565 & -27.76512153 & $3.19361 \pm 0.00088$ & 62.9    & $3.18669 \pm 0.00002$ & 1.4    & MOSFIRE  & $495.5 \pm 63.0$   & $9.43_{-0.10}^{+0.10}$ & $1.387_{-0.10}^{+0.10}$ \\
\nodata          & \nodata & \nodata     & \nodata      & \nodata               & \nodata & $3.18579 \pm 0.00026$ & 18.6   & MUSE ABS & $560.1 \pm 65.7$   & \nodata              & \nodata                \\
1059             & 8203    & 53.15344247 & -27.76611934 & $3.80588 \pm 0.00005$ & 3.1     & $3.80122 \pm 0.00199$ & 124.3  & MUSE ABS & $291.0 \pm 124.4$  & $9.70_{-0.10}^{+0.10}$ & $1.267_{-0.10}^{+0.10}$ \\
1087             & 3506    & 53.16790037 & -27.7979532  & $3.46259 \pm 0.00019$ & 12.8    & $3.45675 \pm 0.00097$ & 65.2   & MUSE ABS & $392.8 \pm 66.6$   & $9.55_{-0.10}^{+0.10}$ & $1.112_{-0.10}^{+0.10}$ \\
1088             & 6012    & 53.15257181 & -27.79384452 & \nodata               & \nodata & $3.08224 \pm 0.00039$ & 28.6   & MUSE ABS & \nodata            & $9.65_{-0.10}^{+0.10}$ & $1.472_{-0.10}^{+0.10}$ \\
1113             & 8528    & 53.16993928 & -27.76833978 & $3.09022 \pm 0.00005$ & 3.7     & $3.08803 \pm 0.00281$ & 206.1  & KMOS     & $160.6 \pm 206.2$  & $9.17_{-0.10}^{+0.17}$ & $0.917_{-0.10}^{+0.10}$ \\
\nodata          & \nodata & \nodata     & \nodata      & \nodata               & \nodata & $3.08520 \pm 0.00065$ & 47.7   & MUSE ABS & $368.4 \pm 47.9$   & \nodata              & \nodata                \\
1138             & 8308    & 53.14850621 & -27.77728375 & $3.61287 \pm 0.00266$ & 172.9   & $3.60520 \pm 0.00047$ & 30.6   & MUSE ABS & $499.3 \pm 175.9$  & $9.40_{-0.10}^{+0.11}$ & $1.567_{-0.10}^{+0.10}$ \\
1215             & 9247    & 53.1487158  & -27.77294489 & $3.33215 \pm 0.00155$ & 107.3   & $3.32623 \pm 0.00104$ & 72.1   & KMOS     & $410.2 \pm 129.4$  & $9.87_{-0.10}^{+0.10}$ & $1.092_{-0.10}^{+0.17}$ \\
\nodata          & \nodata & \nodata     & \nodata      & \nodata               & \nodata & $3.32594 \pm 0.00053$ & 36.7   & MUSE ABS & $430.4 \pm 113.5$  & \nodata              & \nodata                \\
1253             & 8783    & 53.17824938 & -27.77399604 & $3.67218 \pm 0.00096$ & 61.6    & $3.66765 \pm 0.00421$ & 270.4  & KMOS     & $291.0 \pm 277.6$  & $8.59_{-0.10}^{+0.10}$ & $0.587_{-0.10}^{+0.10}$ \\
\nodata          & \nodata & \nodata     & \nodata      & \nodata               & \nodata & $3.66896 \pm 0.00067$ & 43.0   & MUSE EM  & $206.8 \pm 75.2$   & \nodata              & \nodata                \\
1360             & 37765   & 53.17926551 & -27.78289487 & $3.66696 \pm 0.00161$ & 103.4   & $3.66479 \pm 0.00284$ & 182.5  & MUSE ABS & $139.5 \pm 209.9$  & $9.06_{-0.10}^{+0.12}$ & $1.027_{-0.28}^{+0.10}$ \\
1530             & 7002    & 53.17666134 & -27.78380467 & $3.19270 \pm 0.00112$ & 80.1    & $3.18896 \pm 0.00036$ & 25.8   & KMOS     & $267.7 \pm 84.2$   & $8.97_{-0.10}^{+0.10}$ & $0.772_{-0.10}^{+0.10}$ \\
1833             & 3673    & 53.1524905  & -27.79770827 & $3.01437 \pm 0.00178$ & 132.9   & $3.00842 \pm 0.00009$ & 6.7    & MOSFIRE  & $445.0 \pm 133.3$  & $8.49_{-0.10}^{+0.14}$ & $0.362_{-0.36}^{+0.10}$ \\
6375             & 22525   & 53.14339467 & -27.78800227 & $3.41682 \pm 0.00160$ & 108.6   & $3.41287 \pm 0.01914$ & 1300.3 & KMOS     & $268.3 \pm 1306.0$ & $8.79_{-0.10}^{+0.10}$ & $0.882_{-0.10}^{+0.10}$ \\
6518$^{\dagger}$ & 52206   & 53.14325211 & -27.7868279  & $3.75265 \pm 0.00062$ & 39.1    & $3.74969 \pm 0.00020$ & 12.6   & MOSFIRE  & $186.8 \pm 41.1$   &\nodata                 &\nodata \\
6666             & 24954   & 53.15957552 & -27.7767193  & $3.43869 \pm 0.00021$ & 14.2    & $3.43494 \pm 0.00024$ & 16.2   & KMOS     & $253.5 \pm 21.6$   & $9.03_{-0.10}^{+0.10}$ & $0.902_{-0.12}^{+0.10}$ \\
\nodata          & \nodata & \nodata     & \nodata      & \nodata               & \nodata & $3.43492 \pm 0.00008$ & 5.4    & MUSE EM  & $254.8 \pm 15.2$   & \nodata              & \nodata                \\
6878             & 7843    & 53.13953672 & -27.78067557 & $3.60787 \pm 0.00090$ & 58.6    & $3.60020 \pm 0.00064$ & 41.7   & MUSE ABS & $499.8 \pm 72.0$   & $9.40_{-0.11}^{+0.10}$ & $1.387_{-0.13}^{+0.10}$ \\
\nodata          & \nodata & \nodata     & \nodata      & \nodata               & \nodata & $3.60295 \pm 0.00006$ & 3.9    & MOSFIRE  & $320.4 \pm 58.7$   & \nodata              & \nodata                \\
6883             & 9832    & 53.17629722 & -27.77891257 & $3.19503 \pm 0.00222$ & 158.6   & $3.18724 \pm 0.00040$ & 28.6   & MUSE ABS & $557.7 \pm 161.5$  & $9.56_{-0.10}^{+0.10}$ & $1.407_{-0.10}^{+0.10}$ \\
6895             & 5742    & 53.1759691  & -27.79261283 & $3.71273 \pm 0.00194$ & 123.4   & $3.70409 \pm 0.00079$ & 50.3   & MUSE ABS & $550.6 \pm 133.5$  & $8.41_{-0.10}^{+0.10}$ & $0.497_{-0.10}^{+0.10}$ \\
\nodata          & \nodata & \nodata     & \nodata      & \nodata               & \nodata & $3.70522 \pm 0.00020$ & 12.7   & KMOS     & $478.5 \pm 124.3$  & \nodata              & \nodata                \\
8041             & 8635    & 53.1755829  & -27.76874786 & \nodata               & \nodata & $3.69700 \pm 0.00172$ & 109.8  & MUSE ABS & \nodata            & $9.52_{-0.10}^{+0.10}$ & $1.317_{-0.16}^{+0.10}$ \\
8091             & 5468    & 53.14920592 & -27.79147296 & \nodata               & \nodata & $3.55707 \pm 0.00116$ & 76.3   & MUSE ABS & \nodata            & $9.27_{-0.10}^{+0.10}$ & $1.242_{-0.10}^{+0.10}$ \\
8103             & 5741    & 53.17592521 & -27.79246095 & \nodata               & \nodata & $3.70527 \pm 0.00027$ & 17.2   & MUSE EM  & \nodata            & $8.60_{-0.10}^{+0.10}$ & $0.677_{-0.16}^{+0.10}$ \\
\enddata

\tablecomments{(1) MUSE ID (2) \cite{Rafelski2015} ID (3) Right Ascension (4)
  Declination (5) Redshift measured from \Lyalpha\ (the red peak, in the case
  of a double-peaked line) (6) Velocity uncertainty on redshift (7) Systemic
  redshift (8) Velocity uncertainty on systemic redshift (9) Source of systemic
  redshift (MUSE EMission or ABSorption, KMOS, MOSFIRE) (10) \Lya\ velocity
  offset, $\Delta v(\Lya) = c (\zLya - \zsys)/(1+\zsys)$. (11) Stellar mass and
  (12) Star Formation Rate from \textsc{Magphys}, with a minimum uncertainty of
  0.1\,dex.}

\tablenotemark{$\dagger$}{Blended with a $z=0.83$ foreground object.}
\end{deluxetable*}
\end{longrotatetable}

\clearpage
\bibliography{library}{}
\bibliographystyle{aasjournal}

\end{document}